\documentclass[useAMS,usenatbib]{mn2e}
\usepackage{epsf,graphicx,rotating,url,xcolor}

\title[The torus in NGC~5548]{The first spectroscopic dust reverberation programme on active galactic nuclei: the torus in NGC~5548}

\author[H. Landt et al.]{H. Landt$^1$\thanks{E-mail: hermine.landt@durham.ac.uk}\thanks{Visiting Astronomer at the Infrared Telescope Facility, which is operated by the University of Hawaii under contract NNH14CK55B with the National Aeronautics and Space Administration.}, M. J. Ward$^1$\footnotemark[2], D. Kynoch$^1$\footnotemark[2], C. Packham$^{2,3}$, G. J. Ferland$^4$, A. Lawrence$^5$, 
\newauthor
J.-U. Pott$^6$, J. Esser$^6$, K. Horne$^7$, D. A. Starkey$^{7,8}$, D. Malhotra$^7$,  M. M. Fausnaugh$^9$, 
\newauthor
B. M. Peterson$^{10,11,12}$, R. J. Wilman$^1$, R. A. Riffel$^{13}$, T. Storchi-Bergmann$^{14}$, A. J. Barth$^{15}$, 
\newauthor
C. Villforth$^{16}$, H. Winkler$^{17}$ \\ 
$^1$Centre for Extragalactic Astronomy, Department of Physics, Durham University, South Road, Durham, DH1 3LE, UK \\ 
$^2$Department of Physics \& Astronomy, University of Texas at San Antonio, One UTSA Circle, San Antonio, TX 78249, USA \\
$^3$National Astronomical Observatory of Japan, 2-21-1 Osawa, Mitaka, Tokyo 181-8588, Japan \\
$^4$Department of Physics and Astronomy, University of Kentucky, Lexington, KY 40506, USA \\
$^5$Institute for Astronomy, University of Edinburgh, Royal Observatory, Blackford Hill, Edinburgh, EH9 3HJ, UK \\
$^6$Max Planck Institut f\"ur Astronomie, K\"onigstuhl 17, D-69117 Heidelberg, Germany \\
$^7$SUPA Physics and Astronomy, University of St. Andrews, Fife, KY16 9SS, UK \\
$^8$Department of Astronomy, University of Illinois at Urbana-Champaign, Urbana, IL 61801, USA \\
$^9$MIT Kavli Institute for Astrophysics and Space Research, 77 Massachusetts Avenue, 37-241, Cambridge, MA 02139, USA \\
$^{10}$Department of Astronomy, The Ohio State University, 140 West 18th Avenue, Columbus, OH 43210, USA \\
$^{11}$Center for Cosmology and AstroParticle Physics, The Ohio State University, 191 West Woodruff Avenue, Columbus, OH 43210, USA \\
$^{12}$Space Telescope Science Institute, 3700 San Martin Drive, Baltimore, MD 21218, USA \\
$^{13}$Departamento de F\'isica, Centro de Ci\^encias Naturais e Exatas, Universidade Federal de Santa Maria, 97105-900 Santa Maria, RS, Brazil \\
$^{14}$Departamento de Astronomia, Instituto de F\'isica, Universidade Federal do Rio Grande do Sul, CP 15051, 91501-970 Porto Alegre, RS, Brazil \\
$^{15}$Department of Physics and Astronomy, 4129 Frederick Reines Hall, University of California, Irvine, CA, 92697-4575, USA \\
$^{16}$Department of Physics, University of Bath, Claverton Down, Bath BA2 7AY, UK \\
$^{17}$Department of Physics, University of Johannesburg, PO Box 524, 2006 Auckland Park, South Africa}

\begin{document}

\def\la{\mathrel{\hbox{\rlap{\hbox{\lower4pt\hbox{$\sim$}}}\hbox{$<$}}}}
\def\ga{\mathrel{\hbox{\rlap{\hbox{\lower4pt\hbox{$\sim$}}}\hbox{$>$}}}}

\font\sevenrm=cmr7
\def\OIII{[O~{\sevenrm III}]}
\def\FeII{Fe~{\sevenrm II}}
\def\FeIIf{[Fe~{\sevenrm II}]}
\def\SIII{[S~{\sevenrm III}]}
\def\HeI{He~{\sevenrm I}}
\def\HeII{He~{\sevenrm II}}
\def\NeV{[Ne~{\sevenrm V}]}
\def\OIV{[O~{\sevenrm IV}]}

\def\iraf{{\sevenrm IRAF}}
\def\mpfit{{\sevenrm MPFIT}}
\def\galfit{{\sevenrm GALFIT}}
\def\prepspec{{\sevenrm PrepSpec}}
\def\memecho{{\sevenrm MEMEcho}}
\def\mcmcrev{{\sevenrm MCMCRev}}
\def\javelin{{\sevenrm JAVELIN}}
\def\cloudy{{\sevenrm CLOUDY}}
\def\astroimagej{{\sevenrm AstroImageJ}}
\def\banzai{{\sevenrm BANZAI}}

\date{Accepted ~~. Received ~~; in original form ~~}

\pagerange{\pageref{firstpage}--\pageref{lastpage}} \pubyear{2019}

\maketitle

\label{firstpage}

\begin{abstract}

We have recently initiated the first spectroscopic dust reverberation
programme on active galactic nuclei (AGN) in the
near-infrared. Spectroscopy enables measurement of dust properties,
such as flux, temperature and covering factor, with higher precision
than photometry. In particular, it enables measurement of both
luminosity-based dust radii and dust response times. Here we report
results from a one-year campaign on NGC~5548. The hot dust responds to
changes in the irradiating flux with a lag time of $\sim 70$
light-days, similar to what was previously found in photometric
reverberation campaigns. The mean and rms spectra are similar,
implying that the same dust component dominates both the emission and
the variations. The dust lag time is consistent with the
luminosity-based dust radius only if we assume a
wavelength-independent dust emissivity-law, i.e. a blackbody, which is
appropriate for grains of large sizes (of a few $\mu$m). For such
grains the dust temperature is $\sim 1450$~K. Therefore, silicate
grains have most likely evaporated and carbon is the main chemical
component. But the hot dust is not close to its sublimation
temperature, contrary to popular belief. This is further supported by
our observation of temperature variations largely consistent with a
heating/cooling process. Therefore, the inner dust-free region is
enlarged and the dusty torus rather a ``dusty wall'', whose inner
radius is expected to be luminosity-invariant. The dust-destruction
mechanism that enlarges the dust-free region seems to partly affect
also the dusty region. We observe a cyclical decrease in dust mass
with implied dust reformation times of $\sim 5-6$~months.

\end{abstract}


\section{Introduction}

Some active galactic nuclei (AGN) exhibit in their optical spectra
both broad ($\sim 1-5\%$ the speed of light) and narrow emission lines
(type~1 AGN) and some only narrow lines (type~2 AGN). The existence of
these two apparently distinct classes of AGN has been explained within
unified schemes by orientation effects: a geometrically and optically
thick, dusty torus (or warped disc) located outside the accretion disc
obscures the broad emission line region (BLR) for some lines of sight
\citep[see reviews by][]{Law87, Ant93, Urry95, Netzer15}. The
strongest evidence that such a torus exists comes from observations of
broad emission lines in the polarized, scattered light of numerous
type~2 AGN \citep[e.g.,][]{Ant85a, Coh99, Lum01, Tran03} and of strong
infrared (IR) continuum emission at wavelengths $\lambda > 1~\mu$m in
most AGN. The latter is expected since the dusty torus will absorb a
significant fraction of the accretion disc radiation and reradiate it
at longer wavelengths. However, the precise location and geometry of
the dusty torus remain unknown, mainly because it cannot be spatially
resolved. Mid-IR imaging at high angular resolution \citep{Pack05,
  Ramos11}, near-IR and mid-IR interferometric observations
\citep{Tri09, Pott10, Kish11, Bur13, Kish13} and sub-mm
interferometric observations \citep{Garcia16, Imanishi18} of a few of
the brightest AGN have so far imposed mainly upper limits on its
extent (of $\sim 0.1 - 6$~pc).

The most promising technique to constrain the size of the dusty torus
when it cannot be spatially resolved is reverberation, i.e., measuring
the time with which the dust responds to changes in the irradiating
flux. For $\sim 20$ AGN, the radius of the innermost part of the
torus, where dust grains are expected to be close to their sublimation
temperature, has been successfully determined through coordinated,
long-term optical and near-IR photometric campaigns \citep{Clavel89,
  Sit93, Nel96, Okn01, Glass04, Min04, Sug06, Kosh14,
  Vazquez15}. Interestingly, the hot dust radius is variable in most
AGN that have multiple determinations. When photometry can be
performed simultaneously at several near-IR wavelengths, a crude
estimate of the dust temperature is also available. Its variability
behaviour in addition to that of the dust lag times can reveal the
mechanism by which dust is destroyed and reforms and can also
constrain the dust grain type and size \citep{Barv92, Kosh09,
  Schnuelle13, Schnuelle15}. Dust lag times are found to be
systematically smaller than dust radii measured with near-IR
interferometry or estimated from the AGN bolometric luminosity and the
dust sublimation temperature \citep{Okn01, Kish07,
  Nen08a}. \citet{Kaw10, Kaw11} ascribe this to a dust geometry that
is bowl-shaped rather than spherical due to the anisotropy of the
accretion disc emission. Such a geometry would place the hottest dust
further in than the bulk of the emission, thus reducing the observed
dust reverberation times relative to the dust radius estimated from
the continuum radiation. 

There are two approaches that have been taken for the dynamics of the
dusty torus. The hydrostatic scenario depicts the torus as a structure
populated by clouds whose motion is dominated by the gravitational
potential of the central super-massive black hole \citep{Kro88,
  Pier92a, Kro07, Nen08a}. The vertical motion needed to sustain the
clouds in a hydostatic structure with a height-to-radius ratio of
roughly unity could be provided by radiation pressure. The other
scenario advocates the outflow of clouds embedded in a hydromagnetic
disc wind \citep{Koenigl94, Elitzur06, Keating12}. In this approach,
the torus is that region in the wind wherein the clouds are dusty.

With the advent of efficient near-IR cross-dispersed spectrographs
(mounted in particular at 8~m class telescopes) spectroscopic and not
only photometric dust reverberation campaigns are now feasible in
reasonable amounts of observing time. We have recently initiated the
first such programme. The advantages of spectroscopy over photometry
are enormous: (i) a large portion of the hot dust spectral energy
distribution (SED) can be monitored at once and so the temporal
evolution of the dust temperature determined with high accuracy; (ii)
the contribution of the accretion disc to the near-IR continuum flux
can be reliably separated from that of the hot dust, with the former
being a large source of uncertainty in photometry campaigns; (iii)
with accurate dust temperatures and accretion disc luminosities in
hand, luminosity-based dust radii can be determined simultaneously
with the dust lag times and temporal changes in dust covering factor
studied; and (iv) emission lines believed to be associated with the
dusty torus can also be sampled. For example, the coronal
(i.e. forbidden high-ionisation) emission line region, which is
thought to lie betweeen the broad and narrow emission line region
(NELR), although itself dust-free \citep{Ferg97}, is assumed to mark
the onset of the dusty torus. Determining how the line flux and
profile shape vary in relation to the variable hot dust emission can
explain how this region is related to the torus, e.g., through an
X-ray heated wind launched from it \citep{Pier95}. The hot dust
presumably also influences the BLR by limiting its extent
\citep{Netzer93a, Czerny11, Mor12}. Evidence for a limited BLR is
provided by flat-topped \citep{L14} or double-peaked \citep{Storchi17}
line profiles as well as velocity-delay maps (Horne et al., in
preparation). Then, if the BLR is indeed dust- rather than
radiation-bounded, we expect the widths of the profile top to increase
as the hot dust radius decreases. In this case, the BLR might have a
bowl-shaped geometry like the dust \citep{Goad12}.

Here we present our results from a one-year long spectroscopic dust
reverberation campaign on the nearby, well-known AGN NGC~5548
conducted between 2016 August and 2017 July. In this study, we
concentrate on the variability of the hot dust and will discuss the
variability of the coronal lines and broad emission lines elsewhere.
The paper is structured as follows. After we briefly discuss our
science target in Section 2, we give details of the observations, data
reduction and measurements in Section 3. In Sections 4, we derive
response-weighted dust radii. In Section 5, we constrain the location,
dynamics and geometry of the hot dust torus. Finally, in Section 6, we
present a short summary and our conclusions.

\section{The science target}

We have selected the type-1 AGN NGC~5548 as our science target for
several reasons: (i) it has a short and variable dust response time
\citep[$\sim 40 - 80$ days;][]{Kosh14}, which means that in one year
we can measure several dust lags and so can investigate why they vary;
(ii) it has strong and variable near-IR coronal lines \citep{L15b};
(iii) its broad emission line profiles have clearly discernible and
therefore easily separable broad- and narrow-line components, and (iv)
multiple optical reverberation campaigns have determined its black
hole mass \citep[$M_{\rm BH} = 5 \times 10^7~M_\odot$;][]{Pet04,
  Bentz09, Bentz15}. Furthermore, two recent ambitious programs,
namely, the X-ray spectroscopic monitoring campaign in summer 2013 for
$\sim 1.5$~months with {\it XMM-Newton} \citep{Kaastra14, Meh15a} and
the ultraviolet (UV) spectroscopic monitoring campaign with the {\it
  Hubble Space Telescope (HST)} during the first half of 2014
\citep{Storm1, Storm2, Storm3, Storm4, Storm5, Storm6, Storm7}, both
supported by extensive space- and ground-based multiwavelength
observations, make it one of the best-documented AGN yet.

NGC~5548 (J2000 sky coordinates \mbox{R.A. $14^h 17^m 59.5^s$},
\mbox{Decl. $+25^\circ 08' 12''$}) is observable from both the
northern and southern hemispheres. It is at a low redshift
($z=0.01718$), which means that a cross-dispersed near-IR spectrum
samples a large portion of the hot dust component and several strong
near-IR coronal lines. The AGN, which is hosted by a spiral galaxy, is
relatively bright in the near-IR \citep[2MASS $J=11.8$~mag,
  $K_s=10.1$~mag;][]{2MASS}, requiring only modest exposure times on a
4~m class telescope. We adopt here cosmological parameters $H_0 =
70$~km~s$^{-1}$~Mpc$^{-1}$, $\Omega_{\rm M}=0.3$, and
$\Omega_{\Lambda}=0.7$, which give a luminosity distance to NGC~5548
of 74.6 Mpc and an angular scale at the galaxy of 349~pc per arcsec.

\section{The observations}

\subsection{The near-IR spectroscopy} \label{spectroscopy}

\begin{table*}
\caption{\label{obslog} 
IRTF Journal of observations}
\begin{tabular}{lrccrrrrccclc}
\hline
Observation & exposure & airmass & aperture & PA & \multicolumn{3}{c}{continuum S/N} & \multicolumn{2}{c}{telluric standard star} & seeing & cloud & correction \\
Date & (s) && (arcsec$^2$) & ($^{\circ}$) & $J$ & $H$ & $K$ & name & airmass & (arcsec) & condition & factor \\
(1) & (2) & (3) & (4) & (5) & (6) & (7) & (8) & (9) & (10) & (11) & (12) & (13) \\
\hline
2016 Aug 2  & 16$\times$120 & 1.180 & 0.3$\times$3.2 &  96 & 80 & 141 & 220 & HD 131951 & 1.382 & 0.7 & clear   & $0.85\pm0.02$ \\
2016 Aug 11 & 16$\times$120 & 1.209 & 0.3$\times$3.6 &  92 & 66 & 118 & 187 & HD 131951 & 1.425 & 0.6 & photom. & $1.03\pm0.03$ \\
2016 Aug 21 &  5$\times$120 & 1.544 & 0.3$\times$4.0 &  85 & 25 &  48 &  80 & HD 131951 & 1.962 & 0.6 & clear   & $0.71\pm0.06$ \\
2016 Dec 21 & 16$\times$120 & 2.184 & 0.3$\times$4.2 & 281 & 45 &  82 & 133 & HD 131951 & 1.765 & 0.4 & photom. & $0.78\pm0.03$ \\
2017 Jan 6  & 16$\times$120 & 1.355 & 0.3$\times$3.6 & 274 & 46 &  84 & 144 & HD 131951 & 1.178 & 0.9 & clear   & $1.02\pm0.03$ \\
2017 Jan 20 & 16$\times$120 & 1.216 & 0.3$\times$4.2 & 269 & 48 &  84 & 142 & HD 131951 & 1.109 & 0.6 & cirrus  & $0.81\pm0.03$ \\
2017 Jan 24 & 21$\times$120 & 1.240 & 0.3$\times$5.0 & 270 & 39 &  72 & 125 & HD 131951 & 1.078 & 1.2 & clear   & $0.82\pm0.04$ \\
2017 Feb 5  & 16$\times$120 & 1.064 & 0.3$\times$3.4 & 260 & 60 & 108 & 187 & HD 131951 & 1.015 & 0.7 & photom. & $0.67\pm0.02$ \\
2017 Feb 15 & 12$\times$120 & 1.006 & 0.3$\times$4.0 & 248 & 45 &  83 & 142 & HD 131951 & 1.006 & 0.8 & cirrus  & $0.73\pm0.02$ \\
2017 Feb 24 & 16$\times$120 & 1.005 & 0.3$\times$3.2 & 214 & 64 & 109 & 183 & HD 131951 & 1.022 & 0.6 & cirrus  & $1.15\pm0.04$ \\
2017 Mar 17 & 16$\times$120 & 1.027 & 0.3$\times$3.6 & 250 & 48 &  93 & 165 & HD 131951 & 1.005 & 0.7 & photom. & $1.00$ \\
2017 Mar 22 & 18$\times$120 & 1.030 & 0.3$\times$3.4 & 254 & 71 & 133 & 224 & HD 131951 & 1.005 & 0.5 & photom. & $0.77\pm0.02$ \\
2017 May 9  & 16$\times$120 & 1.041 & 0.3$\times$3.6 & 256 & 52 &  94 & 157 & HD 131951 & 1.009 & 0.6 & clear   & $0.89\pm0.03$ \\
2017 May 25 & 16$\times$120 & 1.042 & 0.3$\times$3.2 & 256 & 55 &  99 & 171 & HD 131951 & 1.007 & 0.4 & photom. & $1.11\pm0.04$ \\
2017 Jun 10 & 15$\times$120 & 1.055 & 0.3$\times$3.4 & 258 & 52 &  95 & 162 & HD 121996 & 1.054 & 0.8 & photom. & $0.95\pm0.04$ \\
2017 Jun 20 & 22$\times$120 & 1.049 & 0.3$\times$3.8 & 257 & 54 & 101 & 173 & HD 121996 & 1.043 & 0.7 & cirrus  & $0.80\pm0.03$ \\
2017 Jun 28 & 24$\times$120 & 1.027 & 0.3$\times$3.4 & 251 & 70 & 125 & 211 & HD 121996 & 1.025 & 0.6 & clear   & $0.78\pm0.03$ \\
2017 Jul 3  & 20$\times$120 & 1.005 & 0.3$\times$3.4 & 178 & 80 & 136 & 217 & HD 121996 & 1.001 & 0.5 & photom. & $0.98\pm0.04$ \\
\hline
\end{tabular} 

\parbox[]{18cm}{The columns are: (1) Universal Time (UT) date of
  observation; (2) exposure time; (3) mean airmass; (4) extraction
  aperture; (5) slit position angle, where PA$=0^{\circ}$ corresponds
  to east-west orientation and is defined east through north; S/N in
  the continuum over $\sim 100~\AA$ measured at the central wavelength
  of the (6) J, (7) H, and (8) K bands; for the telluric standard star
  (9) name and (10) mean airmass; (11) seeing in the $K$ band; (12)
  cloud condition and (13) multiplicative photometric correction
  factor relative to the spectrum from 2017 March 17, determined using
  the narrow emission line~\SIII~$\lambda 9531$.}

\end{table*}

We observed the source NGC~5548 between 2016 August and 2017 July
(semesters 2016B and 2017A) with the recently refurbished SpeX
spectrograph \citep{Ray03} at the NASA Infrared Telescope Facility
(IRTF), a 3~m telescope on Maunakea, Hawaii. Our approved proposal
requested a cadence of about a week during the period of a year
starting 2016 August 1. Excluding the 3.5-month period (Sep to
mid-Dec) when the source is unobservable. We were scheduled 24
observing windows of 2-2.5 hours each between 2016 August and 2017
July with an observing gap of about a month (mostly during April) due
to engineering time for the spectrograph. Of the scheduled observing
windows, we lost six due to weather and engineering issues, resulting
in a total of 18 near-IR spectra with an average cadence of about ten
days. We list the journal of observations in Table \ref{obslog}.

We used the short cross-dispersed mode (SXD, $0.7-2.55~\mu$m) equipped
with the $0.3'' \times 15''$ slit, which we oriented at the
parallactic angle. This set-up gives an average spectral resolution of
$R=2000$ or full width at half maximum (FWHM) $\sim 150$~km~s$^{-1}$,
which is sufficiently high to study line profiles and to clearly
discern the narrow and broad line components for the permitted
transitions. The narrow slit also minimizes the flux contamination
from the host galaxy. The on-source exposure time was usually
$16\times120$~s, however, when there was time available we obtained
additional frames. We note that the exposure time for the run on 2016
August 21 was only $5\times120$~s, since we lost most data due to
telescope problems. The chosen exposure time ensured that we obtained
spectra with a high signal-to-noise ($S/N$) ratio in order to reliably
measure emission line profiles. The observations were done in the
usual ABBA nodding pattern. Since the source is extended in the
near-IR, we nodded off onto a blank patch of sky for the background
subtraction.

After the science target, we observed the nearby (in position and
airmass) A0~V star HD~131951 that has accurate optical magnitudes
($B=5.870$~mag, $V=5.901$~mag). We used this standard star to correct
our science spectrum for telluric absorption and for flux
calibration. For the observations in 2017 June and July we used the
A0~V star HD~121996, which is of a similar brightness ($B=5.743$~mag,
$V=5.748$~mag). Flats and arcs were taken mostly after the science
target.

\begin{figure*} 
\centerline{
\includegraphics[scale=0.8, clip=true, bb= 35 235 580 700]{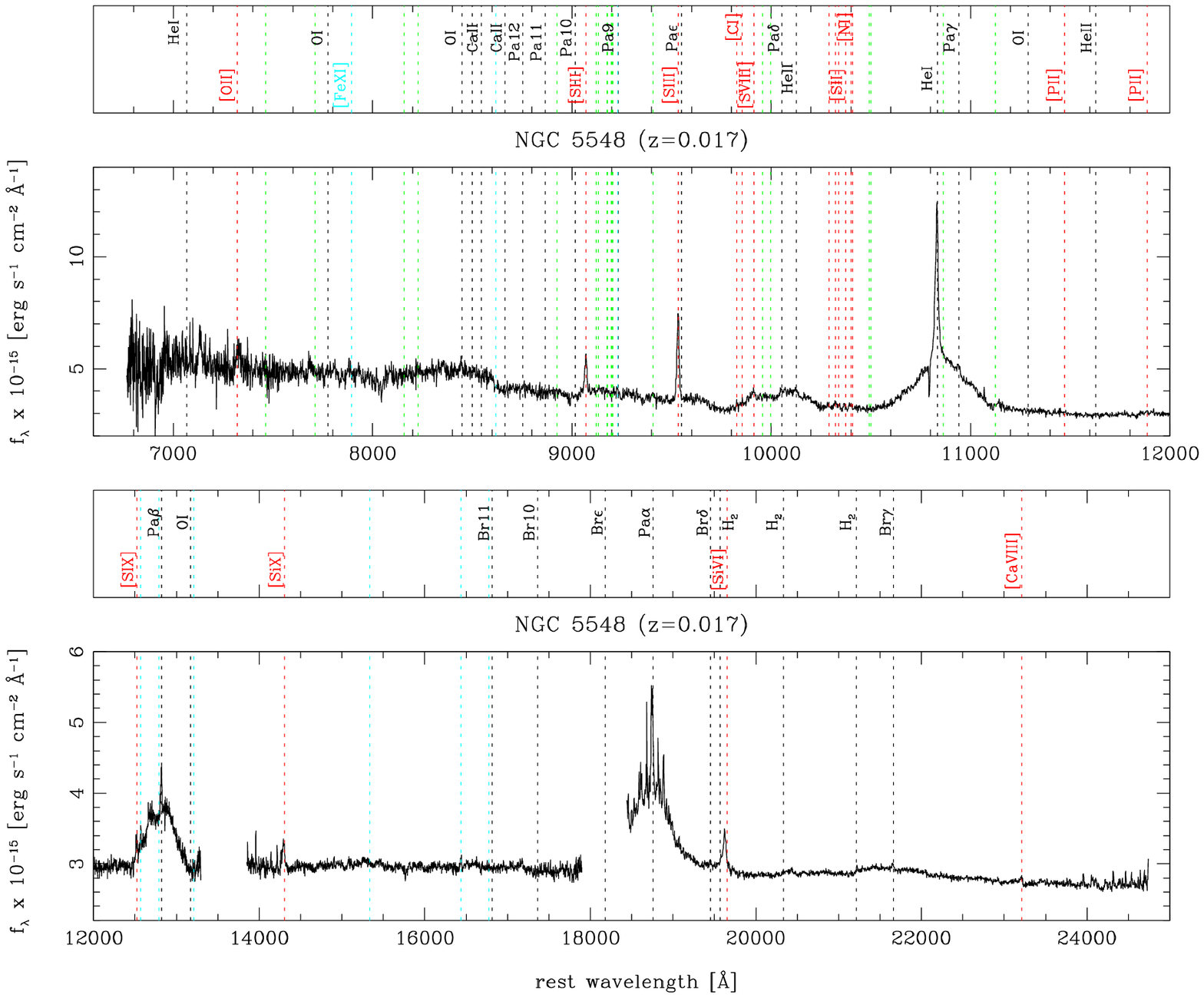}
}
\caption{\label{irtfspec} IRTF SpeX near-IR spectrum from 2017
  February 5 shown as observed flux versus rest-frame
  wavelength. Emission lines listed in Table 4 of \citet{L08a} are
  marked by dotted lines and labeled; black: permitted transitions,
  green: permitted \FeII~multiplets (not labeled), red: forbidden
  transitions and cyan: forbidden transitions of iron (those of
  \FeIIf~not labeled).}
\end{figure*}

We reduced the data using Spextool (version 4.1), an Interactive Data
Language (IDL)-based software package developed for SpeX users
\citep{Cush04}. Spextool carries out all the procedures necessary to
produce fully reduced spectra. This includes preparation of
calibration frames, processing and extraction of spectra from science
frames, wavelength calibration of spectra, telluric correction and
flux-calibration of spectra, and merging of the different orders into
a single, continuous spectrum. We used for both the science target and
the telluric standard star an optimally weighted extraction
\citep{Horne86a}, which requires the fitting and subtracting of a
local background. Based on the Galactic hydrogen column densities
published by \citet{DL90}, the Galactic extinction towards the source
NGC~5548 is negligible and we did not correct for it. In
Fig.~\ref{irtfspec}, we show the spectrum from our run on 2017
February 5 as a representative example.

\subsection{The complementary photometry}

\subsubsection{The GROND observations}

\begin{table*}
\caption{\label{grondphot} 
Optical and near-IR nuclear and host galaxy fluxes from GROND photometry}
\begin{tabular}{lcccccccccccc}
\hline
Observation & \multicolumn{3}{c}{$g'$} & \multicolumn{3}{c}{$r'$} \\
Date & \multicolumn{3}{c}{$\lambda_{\rm eff}=4587$~\AA} & \multicolumn{3}{c}{$\lambda_{\rm eff}=6220$~\AA} \\ 
& nuclear flux & PSF & host galaxy flux & nuclear flux & PSF & host galaxy flux \\
& (erg/s/cm$^2$/\AA) & (arcsec) & (erg/s/cm$^2$/\AA) & (erg/s/cm$^2$/\AA) & (arcsec) & (erg/s/cm$^2$/\AA) \\
\hline
2017 Mar 11 & (6.73$\pm$0.18)e$-$15 & 1.12 & (2.04$\pm$0.07)e$-$14 & (4.79$\pm$0.06)e$-$15 & 1.09 & (2.23$\pm$0.04)e$-$14 \\
2017 Mar 18 & (5.73$\pm$0.15)e$-$15 & 1.23 & (2.15$\pm$0.07)e$-$14 & (4.91$\pm$0.06)e$-$15 & 1.19 & (2.07$\pm$0.04)e$-$14 \\
\hline
Observation & \multicolumn{3}{c}{$i'$} & \multicolumn{3}{c}{$z'$} \\
Date & \multicolumn{3}{c}{$\lambda_{\rm eff}=7641$~\AA} & \multicolumn{3}{c}{$\lambda_{\rm eff}=8999$~\AA} \\
& nuclear flux & PSF & host galaxy flux & nuclear flux & PSF & host galaxy flux \\
& (erg/s/cm$^2$/\AA) & (arcsec) & (erg/s/cm$^2$/\AA) & (erg/s/cm$^2$/\AA) & (arcsec) & (erg/s/cm$^2$/\AA) \\
\hline
2017 Mar 11 & (3.01$\pm$0.01)e$-$15 & 0.92 & (2.70$\pm$0.01)e$-$14 & (2.96$\pm$0.07)e$-$15 & 0.98 & (2.85$\pm$0.07)e$-$14 \\
2017 Mar 18 & (2.55$\pm$0.01)e$-$15 & 1.06 & (2.71$\pm$0.01)e$-$14 & (2.24$\pm$0.05)e$-$15 & 0.98 & (2.16$\pm$0.05)e$-$14 \\
\hline
Observation & \multicolumn{3}{c}{$J$} & \multicolumn{3}{c}{$H$} \\
Date & \multicolumn{3}{c}{$\lambda_{\rm eff}=12399$~\AA} & \multicolumn{3}{c}{$\lambda_{\rm eff}=16468$~\AA} \\
& nuclear flux & PSF & host galaxy flux & nuclear flux & PSF & host galaxy flux \\
& (erg/s/cm$^2$/\AA) & (arcsec) & (erg/s/cm$^2$/\AA) & (erg/s/cm$^2$/\AA) & (arcsec) & (erg/s/cm$^2$/\AA) \\
\hline
2017 Mar 11 & (2.30$\pm$0.01)e$-$15 & 1.45 & (1.57$\pm$0.01)e$-$14 & (2.12$\pm$0.02)e$-$15 & 1.55 & (9.72$\pm$0.08)e$-$15 \\
2017 Mar 18 & (2.17$\pm$0.01)e$-$15 & 1.58 & (1.57$\pm$0.01)e$-$14 & (1.95$\pm$0.02)e$-$15 & 1.87 & (1.14$\pm$0.01)e$-$14 \\
\hline
Observation & \multicolumn{3}{c}{$K$} \\
Date & \multicolumn{3}{c}{$\lambda_{\rm eff}=21706$~\AA} \\
& nuclear flux & PSF & host galaxy flux \\
& (erg/s/cm$^2$/\AA) & (arcsec) & (erg/s/cm$^2$/\AA) \\
\hline
2017 Mar 11 & (1.97$\pm$0.06)e$-$15 & 1.71 & (3.04$\pm$0.16)e$-$15 \\
2017 Mar 18 & (1.93$\pm$0.06)e$-$15 & 1.99 & (2.81$\pm$0.13)e$-$15 \\  
\hline  
\end{tabular} 

\end{table*}

We have obtained simultaneous optical and near-IR photometry in seven
bands with the Gamma-Ray Burst Optical and Near-Infrared Detector
\citep[GROND;][]{grond} mounted on the MPG~2.2~m~telescope, La Silla,
Chile. The observations were taken on 2017 March 11 and 18 with the
latter date only one day after our near-IR spectroscopy observation
from 2017 March 17. The optical channels have a field-of-view (FOV) of
$5.4' \times 5.4'$ and a plate scale of $0.158''$ per pixel, whereas
both the FOV and plate scale of the near-IR channels is larger ($10'
\times 10'$ and $0.6''$ per pixel, respectively). The filter passbands
are relatively wide ($\sim 1000 - 3000$~\AA). The seeing on both
nights was $\sim 0.8''-0.9''$, with clear skies and thin clouds on
2017 March 11 and 18, respectively.

Images were reduced with the \iraf-based pipelines devoloped by
R. Decarli and G. De Rosa \citep{Morg12}, modified and extended for
our purpose. After bias and dark subtraction and flat-fielding of each
individual image, a scaled median sky image constructed from separate
sky exposures was subtracted from each frame. The dither step size was
$18''$ for both the science and sky images, with five dither positions
obtained for each. After sky subtraction, the individual frames were
realigned and combined into the final image. In Table \ref{grondphot},
we list the optical and near-IR nuclear fluxes with their
$1~\sigma$~errors, which were obtained with \galfit~\citep{galfit}, a
software package that models the object's surface brightness profile
with a point spread function (PSF) and a host galaxy component. We
modelled the host galaxy with only a bulge, since the $S/N$ of the
images is not high enough to warrant including also a disc
component. The surface brightness profile of the bulge was
approximated with a S\'ersic profile with S\'ersic index $n=3.0$ and
effective radius $R_{\rm e} = 12.6''$. In addition,
\galfit~simultaneously models the PSF fluxes of the reference stars,
which we used to flux-calibrate the optical and near-IR images based
on the known stellar fluxes as listed in the SDSS \citep{SloanDR7} and
2MASS \citep{2MASS}.

\subsubsection{The LCOGT observations}

\begin{figure}
\centerline{
\includegraphics[scale=0.45]{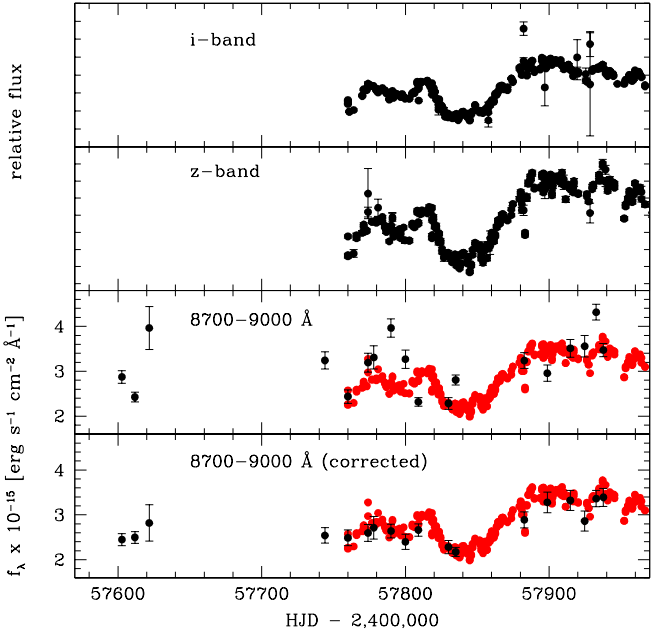}
}
\caption{\label{lcurves} Top two panels: LCOGT $i$ and $z_s$
  light-curves in relative flux units and not corrected for host
  galaxy contribution. Bottom two panels: IRTF SpeX near-IR spectral
  light-curve in the wavelength range $8700-9000$~\AA~(black filled
  circles), both original and corrected using photometric correction
  factors based on the \SIII~$\lambda 9531$~line, versus the $z_s$
  band light-curve corrected for a constant host galaxy contribution
  and scaled to match the IRTF spectral flux from 2017 March 17 (red
  filled circles).}
\end{figure}

We monitored NGC~5548 with the 1~m robotic telescope network of the
Las Cumbres Observatory \citep[LCOGT;][]{LCOGT} as part of the 2014
AGN Key Project almost daily in $V$ between 2016 July 16 and August 8
and in six bands ($V$, $u$, $g$, $r$, $i$ and $z_s$) between 2016
December 12 and 2017 September 9. We make extensive use here of the
light-curves in the $V$ and $z_s$ filters, which have a wavelength
width of 840~\AA~and 1040~\AA~around their central wavelength of
5448~\AA~and 8700~\AA, respectively. The Sinistro cameras used for
imaging have a FOV of $26.5' \times 26.5'$ and a plate scale of
$0.389''$ per pixel. The frames were first processed by LCOGT's
\banzai~pipeline in the usual way (bias and dark subtraction,
flat-fielding and image correction) and were subsequently analysed
with \astroimagej~\citep{astroimagej}. The total flux of the science
target was extracted in a circular aperture with a radius of 15
pixels, from which we subtracted a background flux determined from a
circular region with a minimum and maximum radius of 25 pixels and 35
pixels, respectively. Comparison stars were selected from the same
FOV, whereby we omitted bright, saturated stars and set the number of
stars such that the sum of their flux is $\sim 8-10$~times the flux of
the science target. In Fig.~\ref{lcurves} (top panels), we show the
$i$ and $z_s$ light-curves, which overlap in wavelength with the IRTF
spectrum.

\subsection{The absolute spectral flux scale} \label{rescaling}

\begin{figure}
\centerline{
\includegraphics[clip=true, bb=0 213 577 703, scale=0.45]{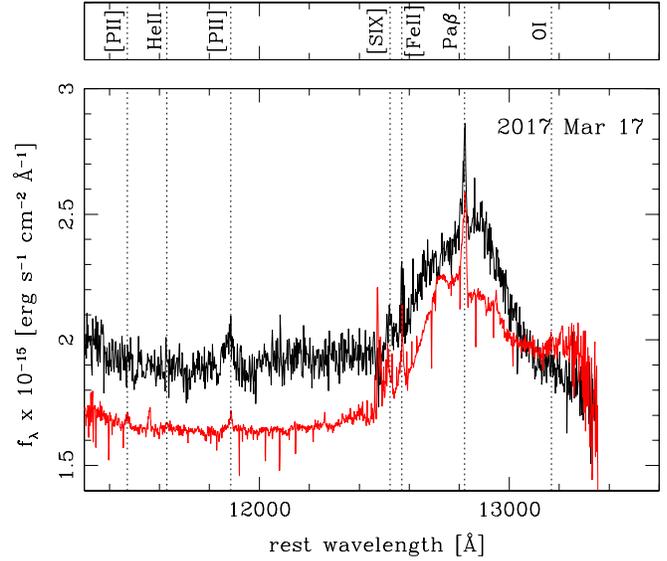}
}
\caption{\label{nifsspec} IRTF SpeX near-IR spectrum from 2017 March
  17 (black) compared to the NIFS $J$ band spectrum obtained by
  summing the flux in a $0.3'' \times 1.2''$ aperture centered on the
  nucleus and oriented at the same position angle as the IRTF spectral
  aperture (red). The main emission lines are identified and labeled.}
\end{figure}

\begin{figure}
\centerline{
\includegraphics[scale=0.45]{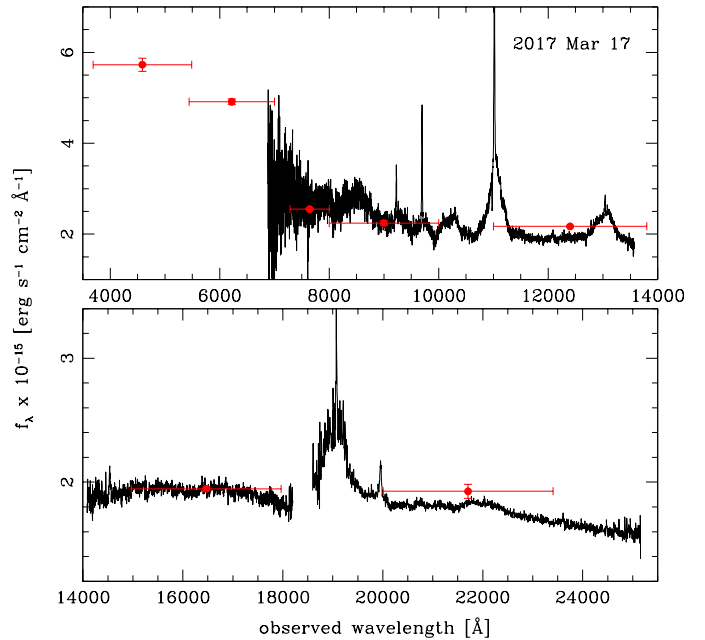}
}
\caption{\label{grondspec} IRTF SpeX near-IR spectrum from 2017 March
  17 (black), overlaid with the optical and near-IR nuclear fluxes
  from the GROND photometry of 2017 March 18 (red filled circles).}
\end{figure}

In order to obtain meaningful light-curves and estimates of phyiscal
parameters such as, e.g. luminosities, it is crucial to know the
absolute flux scale of the spectra as accurately as possible. In
optical spectroscopic reverberation studies, this absolute
spectrophotometry is usually obtained by using the flux of the strong,
narrow emission line \OIII~$\lambda 5007$, which is typically assumed
to stay constant during the campaign and to have a value as measured
in the best-weather spectra. The assumption of constancy is usually
justified for forbidden line transitions, since the emitting gas is
located at pc-to-kpc distances from the ionising source and has a
number density low enough for recombination time-scales to be
large. Then, in order to ensure that the same amount of narrow line
flux is enclosed in all spectra, a wide slit (of a few arcsec) is
used, which is kept at the same position angle \citep[e.g.][]{Bentz08,
  Denney10}. The choice of a high-dispersion grating makes up for the
lost spectral resolution and, due to its relatively narrow wavelength
range, atmospheric dispersion effects can be ignored. The standard
star used for the initial flux calibration is usually observed on the
same night but not necessarily close in time.

The situation is very different in our case. Our near-IR spectra have
a relatively large wavelength range, which we need to study the hot
dust SED. Therefore, we observed with the slit at the parallactic
angle in order to mitigate flux losses due to atmospheric
dispersion. We used the narrowest slit offered ($0.3''$) for two
reasons: to get the spectral resolution necessary to study line shapes
and to minimize the contribution from the host galaxy to the continuum
flux. However, the changing slit position angle and the narrow slit
size mean that we have not necessarily enclosed the same amount of
narrow line flux in all spectra, in particular, if the line-emitting
region is elongated rather than circular or has considerable
small-scale structure like filaments and knots. On the other hand, the
telluric standard star, which we use for the initial flux calibration,
was observed very close in time and airmass to the science
target. Therefore, if slit losses due to seeing for the two were the
same, the spectral flux scale is expected to be accurate even though
the enclosed narrow line flux might vary with slit position angle.

We can best assess if our initial spectral flux scale needs to be
adjusted to reflect an absolute flux scale by using a strong,
non-variable, forbidden narrow emission line in the near-IR as
observed with integral field unit (IFU) spectroscopy. Such
observations are not subject to slit losses since they perform imaging
spectroscopy and so can be used to measure the line flux expected in a
given spectral aperture oriented at a certain angle. The strongest
forbidden narrow emission line in the near-IR is \SIII~$\lambda 9531$,
but there are currently no IFU observations of this line for
NGC~5548. Instead, \citet{Schoenell17} observed NGC~5548 in 2012 at
the Gemini North 8~m telescope with the Near-Infrared Integral Field
Spectrograph \citep[NIFS;][]{nifs} and the Adaptive Optics system
ALTAIR in the $J$ band ($1.14-1.36~\mu$m). These observations,
obtained under a seeing of $0.28''$, which is similar to our IRTF slit
width, and with a FOV of $3'' \times 3''$, cover the weak, forbidden
narrow emission line \FeIIf~$1.2567~\mu$m. They show that the
\FeIIf~emission region is slightly extended (on subarcsec scales) and
has a smooth but elongated structure (see their Fig.~4). For NGC~4151,
there are near-IR IFU observations covering both the \SIII~$\lambda
9531$ and \FeIIf~$1.2567~\mu$m~lines \citep{Storchi09}. These
observations show that the extended morphologies of the two lines are
very similar and so we will assume in the following that any variation
we measure for the \FeIIf~line with changing spectral aperture and/or
orientation angle holds also for the \SIII~line. In any case,
narrow-line reverberation results for NGC~5548 show that the
\OIII~emitting region is 1-3~pc in extent with a density of $\sim
10^5$~cm$^{-3}$ \citep{Pet13}. Since photoionisation models indicate
that all the high-ionisation lines arise in this same component
\citep{Kraemer98}, we expect the \SIII~emission to be dominated by a
nuclear component.

We have re-reduced the NIFS $J$ band \mbox{datacube} for NGC~5548 and
have produced spectra summed within a spectral aperture of $0.3''
\times 1.2''$ centered on the nucleus and oriented at the same
position angle as the IRTF SpeX spectral aperture (see
Fig.~\ref{nifsspec} for an example). The chosen aperture length
ensures that we enclose all the \FeIIf~flux in any given direction
\citep{Schoenell17} and avoids the noisy outer regions of the NIFS
FOV. We have then used the \prepspec~routine developed by K. Horne
\citep[described in][]{Shen15, Shen16} to determine the \FeIIf~flux
variation in the extracted NIFS spectra. \prepspec~models the profiles
of all emission lines, both broad and narrow, and the total
continuum. A time-dependent scaling factor is included to match the
profiles of selected narrow emission lines in all the spectra. As
\citet{vanGron92} and \citet{Faus17} showed, matching the line
profiles rather than directly measuring the line flux gives a much
improved photometric alignment (usually $\la 1-5\%$). The line
profiles are modelled as splines to keep them smooth, but are
otherwise determined by the data. The median photometric correction
factor is held at one, appropriate for a majority of the spectra being
photometric. Using only the \FeIIf~$1.2567~\mu$m~line, we find a
relatively small spread around the median for the different NIFS
spectra, with an average photometric correction factor of only $\sim
2\%$. This is not unexpected since our slit orientation was very
similar for most observations. This then means that we expect the
\FeIIf~line flux, and so the \SIII~line flux, to be constant in our
IRTF spectra. We have then used \prepspec~to determine the spectral
photometric correction factors based on the strong \SIII~$\lambda
9531$~line.

In one case, we can compare the initial spectral flux scale directly
to GROND photometry results (Fig.~\ref{grondspec}). The GROND
observation from 2017 March 18 is only one day after our near-IR
spectrum from 2017 March 17, and so no significant variability is
expected. The GROND nuclear fluxes in the $i'$, $z'$, $J$, $H$ and
$K$~filters, which overlap in wavelength with our near-IR spectrum,
are within $\sim 4\%$, 2\%, 12\%, 0.5\% and 5\%, respectively, of the
spectral fluxes. Note that the much larger value for the $J$ band is
due to two strong, broad emission lines being included in the filter
passband. This high similarity between the GROND nuclear fluxes and
the near-IR spectral fluxes not only confirms that our initial flux
scale for this spectrum, which was observed under photometric weather
conditions, is accurate at the few per cent level, but also strongly
supports our assertion below that the IRTF SpeX spectra are barely
contaminated by host galaxy flux (see Section \ref{components}). We
have then assumed a photometric correction factor of unity for this
spectrum and scaled the photometric correction factors from
\prepspec~for all other spectra relative to it (instead of relative to
the median). The results are listed in Table \ref{obslog}, column
(13). The photometric correction factors lie in the range of $\sim
2-30\%$, with an average of $\sim 15\%$.

In Fig.~\ref{lcurves}, bottom panels, we compare the IRTF spectral
continuum flux in the wavelength range of $8700-9000$~\AA, both
original and corrected, with the $z$ band light-curve from the LCOGT
photometry. We have corrected the $z$ band light-curve for a constant
host galaxy flux contribution, which we estimated using our host
galaxy modelling of the GROND $z'$ band image from 2017 March 18 and
scaled it to match the near-IR spectral flux of the observation from
2017 March 17. Whereas large discrepancies between spectral and
photometry points are evident in about half the cases for the original
data, the two data sets align very well once we have applied the
photometric correction factors based on the \SIII~$\lambda 9531$~line.

Finally, we have also determined the photometric correction factors
inlcuding only the three strongest, permitted narrow emission lines,
namely, \HeI~$1.08~\mu$m, Pa$\beta$, and Pa$\alpha$. Permitted
transitions can be produced in gas of a higher number density and so
can be potentially variable on the time-scales of our monitoring
campaign. We obtain results similar to those using only the
\SIII~$\lambda 9531$~line, which indicates that these lines did not
strongly vary.

\subsection{The spectral continuum components} \label{components}

\begin{figure}
\centerline{
\includegraphics[angle=-90, bb=44 10 600 746, scale=0.32]{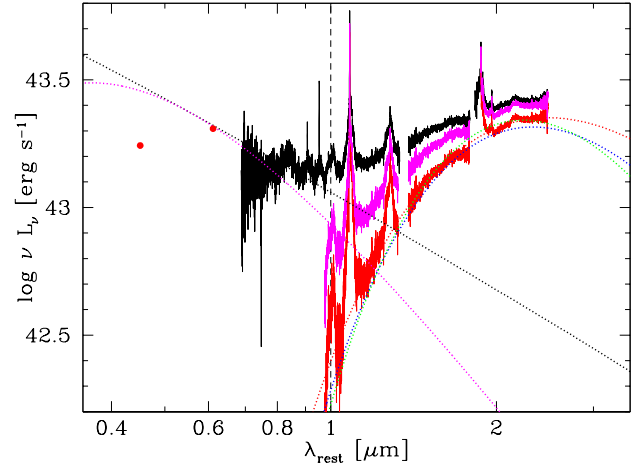}
}
\caption{\label{irsed} IRTF SpeX near-IR spectrum from 2017 March 17
  shown as luminosity versus rest-frame wavelength, together with the
  $g'$ and $r'$ nuclear fluxes from the quasi-simultaneous GROND
  photometry (red circles). We have decomposed the continuum into an
  accretion disc spectrum (with a large outer radius of $r_{\rm
    out}=10^4\,r_{\rm g}$), which approximates the wavelength range of
  $0.7-1~\mu$m (black dotted curve) and still dominates at $1~\mu$m
  (vertical dashed line), and a hot dust component at wavelengths
  $>1~\mu$m (red spectrum). We have fitted the hot dust continuum with
  a blackbody spectrum (red dotted curve) and modified blackbody
  spectra for carbon and for silicate dust (green and blue dotted
  curves, respectively). If the accretion disc cuts off at the
  self-gravity radius ($r_{\rm out} \sim 500\,r_{\rm g}$) or a
  `constant red component' with a blackbody spectrum of temperature $T
  \sim 10^4$~K is assumed to dominate the emission at wavelengths $\la
  1~\mu$m (magenta dotted curve), the hot dust spectrum is broader
  (magenta spectrum).}
\end{figure}

\begin{table*}
\caption{\label{lumradiustab} 
Physical parameters for the calculation of luminosity-weighted dust radii}
\begin{tabular}{lclcclcclcc}
\hline
Observation & accretion & \multicolumn{3}{c}{blackbody} & \multicolumn{3}{c}{silicate dust} & \multicolumn{3}{c}{carbon dust} \\
Date & disc & \multicolumn{3}{c}{($\beta=0$)} & \multicolumn{3}{c}{($\beta=-1$)} & \multicolumn{3}{c}{($\beta=-2$)} \\
& log~$L_{\rm uv}$ & $T_{\rm d}$ & log~$L_{\rm d}$ & $R_{\rm d,lum}$ & $T_{\rm d}$ & log~$L_{\rm d}$ & $R_{\rm d,lum}$ & $T_{\rm d}$ & log~$L_{\rm d}$ & $R_{\rm d,lum}$ \\
& (erg/s) & (K) & (erg/s) & (lt-days) & (K) & (erg/s) & (lt-days) & (K) & (erg/s) & (lt-days) \\
(1) & (2) & (3) & (4) & (5) & (6) & (7) & (8) & (9) & (10) & (11) \\
\hline
2016 Aug 2  & 44.44 & 1432$\pm$12 & 44.11 & 59 & 1212$\pm$9  & 43.12 & 564 & 1052$\pm$6  & 44.04 & 367 \\
2016 Aug 11 & 44.44 & 1467$\pm$14 & 44.10 & 56 & 1240$\pm$10 & 43.23 & 539 & 1075$\pm$7  & 44.00 & 351 \\
2016 Aug 21 & 44.55 & 1458$\pm$29 & 44.15 & 64 & 1231$\pm$21 & 43.14 & 620 & 1067$\pm$15 & 44.02 & 404 \\
2016 Dec 21 & 44.46 & 1444$\pm$20 & 43.97 & 59 & 1219$\pm$14 & 43.10 & 570 & 1056$\pm$10 & 43.76 & 372 \\
2017 Jan 6  & 44.43 & 1430$\pm$16 & 44.00 & 58 & 1207$\pm$12 & 43.09 & 562 & 1046$\pm$9  & 43.93 & 367 \\
2017 Jan 20 & 44.49 & 1432$\pm$20 & 43.97 & 62 & 1211$\pm$14 & 43.09 & 598 & 1050$\pm$11 & 43.94 & 390 \\
2017 Jan 24 & 44.50 & 1466$\pm$25 & 43.93 & 60 & 1232$\pm$17 & 43.07 & 585 & 1063$\pm$13 & 43.97 & 385 \\
2017 Feb 5  & 44.45 & 1390$\pm$13 & 44.03 & 63 & 1180$\pm$10 & 43.10 & 602 & 1026$\pm$8  & 43.61 & 390 \\
2017 Feb 15 & 44.41 & 1411$\pm$17 & 43.98 & 58 & 1196$\pm$12 & 43.08 & 559 & 1039$\pm$9  & 43.68 & 363 \\
2017 Feb 24 & 44.48 & 1396$\pm$15 & 43.95 & 64 & 1188$\pm$11 & 43.04 & 615 & 1035$\pm$8  & 43.67 & 397 \\
2017 Mar 17 & 44.36 & 1478$\pm$17 & 43.98 & 50 & 1245$\pm$12 & 43.10 & 487 & 1076$\pm$9  & 43.58 & 320 \\
2017 Mar 22 & 44.36 & 1453$\pm$14 & 44.03 & 52 & 1230$\pm$9  & 43.13 & 499 & 1067$\pm$7  & 43.91 & 325 \\
2017 May 9  & 44.54 & 1471$\pm$18 & 43.96 & 62 & 1238$\pm$13 & 43.04 & 606 & 1070$\pm$9  & 43.91 & 398 \\
2017 May 25 & 44.60 & 1439$\pm$14 & 44.04 & 70 & 1215$\pm$10 & 43.10 & 675 & 1053$\pm$7  & 43.70 & 440 \\
2017 Jun 10 & 44.60 & 1515$\pm$19 & 43.99 & 63 & 1270$\pm$13 & 43.13 & 617 & 1094$\pm$10 & 43.98 & 408 \\
2017 Jun 20 & 44.54 & 1477$\pm$17 & 44.02 & 62 & 1245$\pm$11 & 43.11 & 600 & 1077$\pm$8  & 43.94 & 392 \\
2017 Jun 28 & 44.63 & 1474$\pm$13 & 44.02 & 69 & 1243$\pm$9  & 43.12 & 667 & 1076$\pm$7  & 43.60 & 436 \\
2017 Jul 3  & 44.63 & 1474$\pm$13 & 44.08 & 69 & 1248$\pm$9  & 43.17 & 662 & 1083$\pm$7  & 43.62 & 430 \\
\hline
\end{tabular} 

\parbox[]{16.5cm}{The columns are: (1) Universal Time (UT) date of
  observation; (2) total accretion disc luminosity; for a blackbody
  emissivity (3) dust temperature; (4) total dust luminosity and (5)
  dust radius; for an emissivity law appropriate for silicate dust
  with small grain sizes of $a \la 0.1~\mu$m (6) dust temperature; (7)
  total dust luminosity and (8) dust radius; for an emissivity law
  appropriate for carbon dust with small grain sizes of $a \la
  0.1~\mu$m (9) dust temperature; (10) total dust luminosity and (11)
  dust radius.}

\end{table*}

In this study, we intend to measure dust lag times and determine
luminosity-based dust radii in order to compare the two. For the first
we need to assemble the light-curves of both the irradiating
(accretion disc) flux and hot dust emission, for the latter we need to
measure the dust temperature and estimate the UV/optical accretion
disc luminosity responsible for heating the dust. All of this is
possible using our cross-dispersed near-IR spectra. Their large
wavelength range covers about half of the extent of the hot dust SED,
which means that we determine the dust temperature with high precision
and measure the dust flux near its peak, where it is least
contaminated by other emission components. Then, as \citet{L11b, L11a}
showed, the accretion disc still dominates the continuum flux at $\sim
1~\mu$m. So with the increased wavelength coverage in the blue of the
refurbished SpeX (as short as $\sim 0.7$~$\mu$m) we sample a
considerable part of this component, simultaneously with the hot dust
SED.

Our near-IR spectral aperture also includes flux from the host
galaxy. This flux, however, constitutes in this low-redshift source
the entire local background and so is largely removed from the final
spectrum during the data reduction process (see Section
\ref{spectroscopy}). This assertion is supported by the close
similarity between the GROND PSF fluxes and near-IR spectral fluxes
for the observation of 2017 March 17 (see Section \ref{rescaling} and
Fig.~\ref{grondspec}). As a further test, we have, on the one hand,
estimated for this observation the host galaxy flux enclosed in the
near-IR spectral aperture using the correlation of \citet{L11b}
between spectral aperture and host galaxy flux based on {\it Hubble
  Space Telescope} images and, on the other hand, repeated the data
reduction process without subtracting a local background. Both
approaches give a similar result, indicating that the spectral flux in
the $J$ band, where the host galaxy spectrum has its maximum
\citep[see e.g. Fig.~6 of][]{L11b}, would be $\sim 20-30\%$~higher if
the host galaxy flux was still included in the final spectrum.

In order to determine the dust temperature and luminosity and to
estimate the UV/optical accretion disc luminosity, we have decomposed
the spectral continuum into its two components. For this purpose, we
have first approximated the rest-frame wavelength range of
$0.7-1~\mu$m with an accretion disc spectrum, which we have
subsequently subtracted from the total spectrum. We have then fitted
the resultant hot dust spectrum at wavelengths $>1~\mu$m with a
blackbody, representing emission by large dust grains, and with two
modified blackbodies, approximating the emissivity of sub-micron
silicate and carbon dust grains. Fig.~\ref{irsed} shows as an example
the observation from 2017 March 17 for which the quasi-simultaneous
GROND photometry extends the near-IR spectral range into the
optical. Table \ref{lumradiustab} lists the relevant physical
parameters extracted from the spectral decomposition.

\subsubsection{The accretion disc spectrum} \label{accspectrum}

\begin{figure}
\centerline{
\includegraphics[angle=-90, scale=0.32]{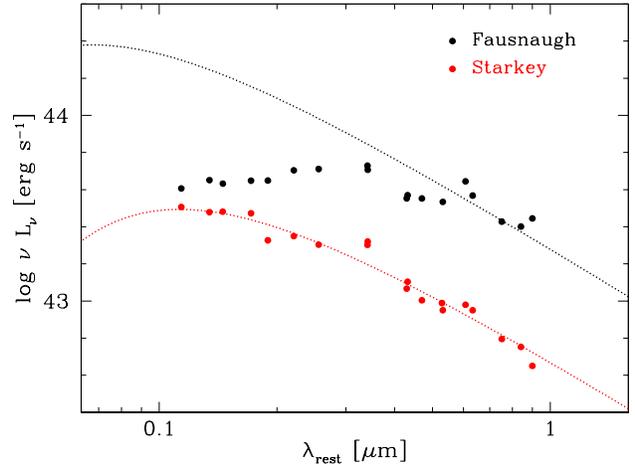}
}
\caption{\label{mehdipour} Photometric data from the 2014 {\it HST}
  reverberation campaign on NGC~5548. The mean observed AGN spectrum
  presented by \citet{Storm3} (black filled circles), corrected for
  host galaxy starlight and line emission, is much redder than the
  prediction from a standard accretion disc with an accretion rate of
  $\sim 0.1~M_\odot$ per year (black dotted line). However, the mean
  accretion disc spectrum inferred by \citet{Storm6} from fits to the
  observed light-curves (red filled circles) can be well approximated
  with a standard accretion disc, if a relatively low accretion rate
  of $\sim 0.01~M_\odot$ per year is assumed (red dotted line).}
\end{figure}

\begin{figure}
\centerline{
\includegraphics[angle=-90, scale=0.32]{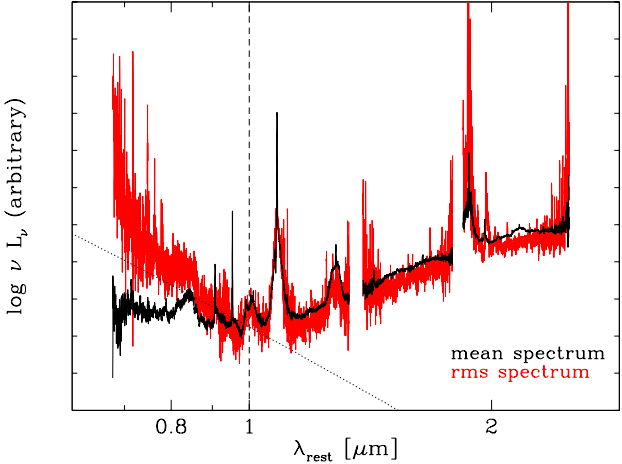}
}
\caption{\label{rmsspec} The mean (black) and variable (rms) spectrum
  (red) for our campaign normalised at rest-frame $1~\mu$m (vertical
  dashed line). The spectrum of a standard accretion disc (black
  dotted line) approximates well the wavelength range of $0.7-1~\mu$m
  for the variable component, which is observed to be much bluer than
  the mean spectrum (similar to the result from the 2014 {\it HST}
  reverberation campaign; see Fig.~\ref{mehdipour}). The spectral
  shape at wavelengths $\ga 1~\mu$m is similar in the mean and rms
  spectrum.}
\end{figure}

We calculated the accretion disc spectrum for a steady, geometrically
thin, optically thick accretion disc, in which case the emitted flux
is independent of viscosity and each element of the disc face radiates
roughly as a blackbody with a characteristic temperature depending
only on the mass of the black hole, $M_{\rm BH}$, the accretion rate,
$\dot{M}$, and the radius of the innermost stable orbit
\citep[e.g.,][]{Peterson, FKR}. We adopted the Schwarzschild geometry
(non-rotating black hole) and for this the innermost stable orbit is
at $r_{\rm in}=6\,r_{\rm g}$, where $r_{\rm g}=G M_{\rm BH}/c^2$ is
the gravitational radius, with $G$ the gravitational constant and $c$
the speed of light. Furthermore, we have assumed that the disc is
viewed face-on and that it extends to $r_{\rm out}=10^4 r_{\rm
  g}$. With these assumptions, only the mass and accretion rate of the
black hole are additionally required in order to constrain the
accretion disc spectrum. The black hole mass of NGC~5548 has been
derived by numerous optical reverberation campaigns and is estimated
to be $M_{\rm BH}=(5.2 \pm 0.2) \times 10^7~M_\odot$ \citep{Pet04,
  Bentz09, Bentz15}, assuming a geometrical scaling factor of $f=4.3$
\citep{Grier13b} to convert the measured virial product to black hole
mass. This black hole mass is uncertain by $\sim 0.4$~dex due to the
intrinsic scatter of $f$~values for individual AGN, but is consistent
with the value of $M_{\rm BH}=(3.9^{+2.9}_{-1.5}) \times 10^7~M_\odot$
derived by \citet{Pan14} based on dynamical modelling and the
preliminary value of $\sim (7 \pm 2) \times 10^7~M_\odot$ (Horne et
al., in preparation) derived from the velocity-delay map of the
2014~{\it HST}~campaign data, both of which do not require a scaling
factor. The accretion rate can be obtained directly from an
approximation of the accretion disc spectrum to the data. In Table
\ref{lumradiustab}, column (2), we list the resultant total accretion
disc luminosity, $L_{\rm uv}$, which we have used in Section
\ref{lumradius} to calculate the luminosity-based dust radius. We note
that our approach attributes the observed changes in accretion disc
luminosity to changes in accretion rate, although on the time-scales
of our monitoring campaign the accretion disc variability is almost
certainly due to reprocessing of the variable high-energy radiation
(see also Section \ref{reverberation}). However, the total accretion
disc spectrum in the low energy range covered by our near-IR spectrum
does not change substantially if reprocessed flux is added, and by
omitting this component we expect to underestimate the total accretion
disc luminosity by only $\sim 10-20\%$ \citep[][see their
  Fig.~3]{Gardner17}.

The spectrum of a standard accretion disc such as the one assumed here
is observed in AGN for the variable \citep[e.g.][]{Collier99} and
polarized components of the optical/near-IR continuum \citep{Kish08}
and also for the total flux in AGN with negigible contribution from
host galaxy starlight \citep[e.g.][]{Korat99, L11a}. Therefore, we are
confident that our spectral decomposition reliably isolates the hot
dust component. However, the 2014 {\it HST} reverberation campaign on
NGC~5548 showed that, whereas the mean accretion disc spectrum as
inferred from the light-curves by \citet{Storm6} is compatible with
the standard accretion disc spectrum, albeit assuming a relatively low
accretion rate, the mean observed UV/optical AGN continuum fluxes
(corrected for host galaxy and line emission) presented by
\citet{Storm3} lie well below the extrapolation of the standard
accretion disc spectrum to shorter wavelengths
(Fig.~\ref{mehdipour}). This spectral difference is also evident in
our data. In Fig.~\ref{rmsspec}, we show the mean and variable (rms)
spectra for our campaign, which we have calculated following
\citet{Pet04}. Whereas the mean spectrum at rest-wavelengths $\la
1~\mu$m is redder than the spectrum of a standard accretion disc, this
model approximates well the rms spectrum in this wavlength range. In
this respect, see also the example in Fig.~\ref{irsed}, which shows
the GROND $r'$ band photometry to agree well with the calculated
accretion disc spectrum but the observed $g'$ band flux to lie well
below it. The flux difference between the mean spectrum of
\citet{Storm3} and the accretion disc spectrum of \citet{Storm6}
indicates the presence of a `constant red component'. This component
has roughly the spectrum of a blackbody with a temperature of $T\sim
10^4$~K and, therefore, could be the reprocessor predicted by
\citet{Gardner17}.

The presence of the `constant red component' means that our accretion
disc spectrum overestimates the UV/optical luminosity available to
heat the dust, and so the dust radii based on it. Since in the two
interpretations of the data from the 2014 {\it HST} reverberation
campaign the total accretion disc luminosities differ by a factor of
$\sim 10$, we estimate that we may have overestimated the hot dust
radius by up to a factor of $\sim 3$. The dominance of the `constant
red component' at the shortest wavelengths of our near-IR spectrum
could also mean that it is its spectral shape rather than that of the
accretion disc that determines the resultant dust spectrum. If we
assume this, more flux at short wavelengths is attributed to dust than
in the case of the accretion disc, thus broadening the dust spectrum
(Fig.~\ref{irsed}). However, this scenario is unlikely. Since the
shape of the variable (rms) and mean spectrum are similar at
wavelengths $\ga 1~\mu$m (Fig.~\ref{rmsspec}), only the variable
components (accretion disc and hot dust) contribute to the near-IR
flux.

The spectrum of the isolated hot dust component (Section
\ref{lumradius}) strongly depends on the assumed accretion disc
spectrum at long wavelengths, which in turn is sensitive to the chosen
outer radius of the accretion disc. We have assumed a very large outer
radius, which gives a maximum contribution from the accretion disc at
near-IR wavelengths. In order for the accretion disc to dominate the
total flux at $\sim 1~\mu$m, as established by the near-IR
radius-luminosity relationship \citep{L11b}, the outer radius can be
as small as $r_{\rm out} \sim 1500\,r_{\rm g} \sim
4.4$~light-days. Choosing instead this smaller value, does not alter
significantly the resultant hot dust spectrum. Accretion discs are
assumed to become unstable and fragment if their extent considerably
exceeds the self-gravity radius. In our case, this value is reached at
$r_{\rm out} \sim 500\,r_{\rm g} \sim 1.5$~light-days. If we assume
this small outer radius, the accretion disc spectrum at near-IR
wavelengths is very similar to that of the `constant red component',
thus resulting in a similarly broad dust spectrum
(Fig.~\ref{irsed}). We further consider this case below and show that
such a broad dust spectrum is unlikely, since it implies a large
temperature range and thus that the hottest dust is much more variable
than the coldest dust. This would change the spectral shape of the
variable (rms) spectrum relative to that of the mean spectrum, which
is contrary to what we observe (Fig.~\ref{rmsspec}). An unexpectedly
large accretion disc radius for NGC~5548 was also inferred by
\citet{Storm2} based on the 2014 {\it HST} reverberation campaign
data. They found a lag time of $\tau=0.35$~light-days at 1367~\AA,
which translates to a radius of $r_{\rm out} \ga 5$~light-days for
wavelengths $\lambda \ga 1~\mu$m using the standard relationship of
$\tau \propto \lambda^{4/3}$.

\subsubsection{The hot dust spectrum} \label{lumradius}

\begin{figure}
\centerline{
\includegraphics[scale=0.42]{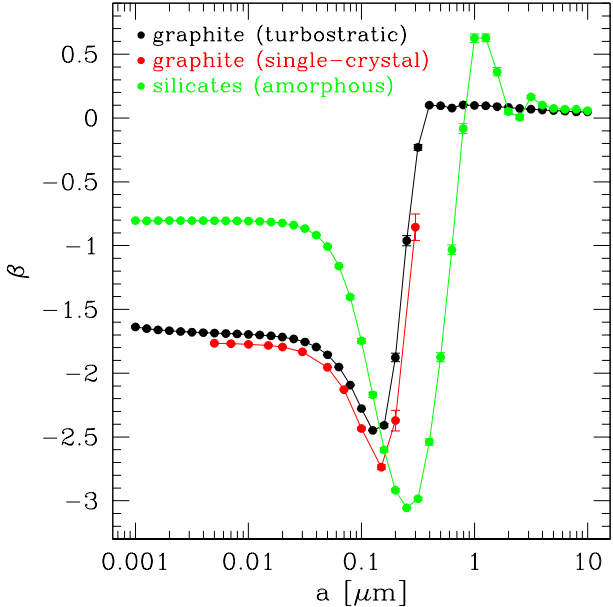}
}
\caption{\label{qext} Power-law slope $\beta$ versus the grain size
  for turbostratic graphite (black), single-crystal graphite (red) and
  amorphous silicates (green). We have fitted the relationship
  $Q_{\lambda} (a) \propto \lambda^{\beta}$ in the wavelength region
  of $\lambda=1-3~\mu$m the data calculated by \citet{Draine16} for
  graphite and by \citet{Draine17} for silicates.}
\end{figure}

Assuming that the hot dust is optically thick to the incident
UV/optical radiation but optically thin to its own thermal radiation,
its emission depends on the temperature, $T$, and the wavelength,
$\lambda$, as:

\begin{equation}
\label{dustflux}
F_{\rm d} (\lambda) = \pi B_\lambda (T) Q_{\lambda} (a),
\end{equation}

\noindent 
where $B_\lambda (T)$ is the Planck function and $Q_{\lambda} (a)$ is
the emission efficiency of the dust at IR wavelengths, which depends
on the grain size $a$. Assuming radiative equilibrium, the dependence
of the dust temperature on the luminosity of the irradiating source
can then be derived by equating the flux absorbed by a spherical dust
grain to that radiated by it:

\begin{equation}
\label{dusteq}
\frac{\pi a^2}{4 \pi R_{\rm d,lum}^2} \int L_{\lambda} Q_{\lambda}^{\rm abs} d\lambda = 4 \pi a^2 \int F_{\rm d} (\lambda) d\lambda,
\end{equation}

\noindent
where $R_{\rm d,lum}$ is the luminosity-weighted dust radius and
$Q_{\lambda}^{\rm abs}$ is the absorption efficiency of the dust in
the UV/optical. We assume that $Q_{\lambda}^{\rm abs}=1$, i.e. that
the absorption cross section equals the geometrical cross section
\citep[e.g][]{Rees69} and Planck-average to get:

\begin{equation}
\label{Stefan-Boltz}
\frac{L_{\rm uv}}{4 \pi R_{\rm d,lum}^2} = 4 \sigma T^4 \langle Q^{\rm em} \rangle,
\end{equation}

\noindent
where $\sigma$ is the Stefan-Boltzmann constant and $\langle Q^{\rm
  em} \rangle$ is the Planck-averaged value of $Q_{\lambda} (a)$. We
consider here only a single grain size, whereas in general the dust
will have a grain size distribution. However, it can be shown that
this is a reasonable approximation for our limited near-IR wavelength
range if it samples a single temperature, since the emission is
dominated by the largest and hottest grains.

We have subtracted from the total near-IR spectrum that of the
accretion disc and have fitted eq. \ref{dustflux} to the resulting
dust spectrum at wavelengths $>1~\mu$m (Fig.~\ref{irsed}). For this we
have assumed that the IR emissivity can be approximated with a
power-law of the form $Q_{\lambda} (a) \propto \lambda^{\beta}$. We
have used the extinction efficiencies recently calculated by B. Draine
and collaborators to determine the power-law slope $\beta$ in the
wavelength range of interest ($\lambda = 1-3~\mu$m) for different
grain sizes ($a=0.001 - 10~\mu$m) for (turbostratic and
single-crystal) graphite \citep{Draine16} and (amorphous) silicates
\citep{Draine17}. The results are shown in Fig.~\ref{qext}. Both dust
types span a relatively large range in $\beta$, however, the
assumption of $\beta=-2$ and $\beta=-1$ is appropriate for carbon and
silicate dust, respectively, for small grain sizes of $a \la
0.1~\mu$m. The blackbody case ($\beta=0$) is reached for grain sizes
of $a \ga 0.4~\mu$m and $\ga 2~\mu$m for carbon and silicate dust,
respectively. We have performed the fits with the C routine
\mbox{\mpfit} \citep[version 1.3a;][]{mpfit}, which solves the
least-squares problem with the Levenberg-Marquardt technique, and have
fitted for the temperature and flux scaling assuming three emissivity
laws ($\beta=0$, $-1$ and $-2$). We have included in the fit only the
continuum part of the near-IR spectrum, i.e., we have excluded
emission lines, and have rebinned it to \mbox{$\Delta \log \nu =
  0.01$~Hz}. The errors were calculated in the usual way as the
standard error on the binned flux average. We note that, since about
half the hot dust spectrum is constrained by the data, its temperature
and flux are well defined. We obtain average temperatures of
\mbox{$\langle T \rangle = 1450\pm7$~K}, \mbox{$1225\pm5$~K} and
\mbox{$1061\pm4$~K} for an emissivity law with $\beta=0$, $-1$ and
$-2$, respectively.

We have then calculated luminosity-weighted dust radii using the
best-fit dust temperatures and approximated $L_{\rm uv}$ with the
accretion disc luminosity. For the Planck-averaged emission
efficiencies we used in the case of \mbox{$\beta=-1$} a value of
\mbox{$\langle Q^{\rm em} \rangle=0.0210$} appropriate for silicates
of \mbox{$T=1259$~K} and \mbox{$a=0.1~\mu$m} \citep{Laor93} and in the
case of \mbox{$\beta=-2$} a value of \mbox{$\langle Q^{\rm em}
  \rangle=0.0875$} appropriate for graphite of \mbox{$T=1000$~K} and
\mbox{$a=0.1~\mu$m} \citep{Draine16}\footnote{The calculated data are
  provided by B. Draine in machine readable form at
  \url{http://www.astro.princeton.edu/~draine/dust/}}. We obtain
average luminosity-weighted dust radii of \mbox{$\langle R_{\rm d,lum}
  \rangle = 61\pm1$~light-days}, \mbox{$385\pm8$~light-days} and
\mbox{$590\pm12$~light-days} in the case of a blackbody, and
small-grain carbon and silicate dust, respectively. Since for the
latter we have considered only solutions for \mbox{$a=0.1~\mu$m}, the
resulting hot dust radius will be larger by up to a factor of $\sim
10$ for grain sizes as small as \mbox{$a=0.001~\mu$m}. In addition, we
might have overestimated in all three cases the total accretion disc
luminosity by up to a factor of $\sim 10$ (see Section
\ref{accspectrum}), resulting in an overestimation of the hot dust
radius by a factor of $\sim 3$.

The unknown outer radius of the accretion disc introduces an
uncertainty in the shape of the hot dust spectrum. As discussed in
Section \ref{accspectrum}, assuming that the disc cuts off already at
the self-gravity radius, gives a much broader dust spectrum. Such a
broad spectrum is much wider than a single blackbody
(Fig.~\ref{irsed}). Fitting the sum of two blackbodies to this broad
dust spectrum for the observation from 2017 March 17, we obtain
best-fit temperatures of \mbox{$T \sim 1800$~K} and \mbox{$\sim
  960$~K}, which results in dust radii of \mbox{$R_{\rm d,lum} \sim
  35$~light-days} and $\sim 120$~light-days for the two
components. This large difference in radius is expected to affect the
observed variability, which should be dominated by the fastest
variations and so by the hottest component. Therefore, in this
scenario, we expect the spectral shape of the variable (rms) spectrum
to be dominated by the hottest dust component and so to be
significantly different from that of the mean spectrum. This is
contrary to what we observe; the spectral shape at wavelengths $\ga
1~\mu$m in the rms spectrum is similar to that in the mean spectrum
(Fig.~\ref{rmsspec}), indicating that the hot dust region spans a very
narrow range in radius thus mimicking a single dust component. A
narrow range in dust radius is also suggested by our reverberation
results, which do not show any significant difference between the $H$
and $K$ bands (Section \ref{revresults}).

\section{Response-weighted dust radii} \label{reverberation}

It is currently assumed that the variability of the hot dust is driven
by the variability of the accretion disc, since this provides the
UV/optical photons necessary to heat the dust grains. In turn, the
emission variability of the accretion disc flux on time-scales less
than a year is attributed to reprocessing of high-energy radiation
from an intrinsically variable source located above the accretion disc
and illuminating it (the lamp-post model). Assuming that the dust
absorbs and re-emits instantaneously the incident UV/optical photon
flux, we can estimate the dust radius based on light-travel time
arguments alone as $R_{\rm d,rev} = \tau \, c$, where $\tau$ is the
reverberation lag time, i.e. the time between the emission of the
accretion disc flux and the re-emission of this flux by the dust, with
the lag in the observer's frame increased by a factor of $1+z$. In
practice, the dust will have an extended geometry and so an emissivity
distribution that has a lag distribution (also referred to as delay
map, transfer function or response function) associated with it.

The difficulty lies in deriving the delay map from the observed
light-curves. The echo light-curve is just a smoothed and delayed
version of the driving light-curve, i.e. the driving light-curve
shifted by a lag and convolved with the geometry-dependent lag
distribution \citep{Bla82b, Pet93}. With very high-quality data,
i.e. extremely well-sampled light-curves, the observed transfer
function can be recovered by Fourier transforming in time, and so
knowledge about the geometry and physical conditions of the
reprocessor gained. If the data are of relatively poor-quality,
i.e. irregularly and sparsely sampled light-curves, the observed
transfer function can be recovered by other methods, e.g. the maximum
entropy method \citep{Horne94} and Markov chain Monte Carlo (MCMC)
sampling of model parameters \citep{Pan11, Starkey16}, or just the
mean lag time estimated using, e.g., cross-correlation methods
\citep{Gas87, White94, Zu11}.

\citet{Kosh14} used cross-correlation analysis to estimate the mean
hot dust response time in NGC~5548 for six monitoring periods and
found values varying in the range of $\sim 40 - 80$~days. This result
is interesting since it may imply that the hot dust distribution is
not constant with time but changes, possibly due to dust grain
destruction and reformation processes \citep{Kosh09}. The length of
our near-IR monitoring campaign is 336~days and so spans $\sim 4 -
8$~times the mean reverberation lag time. However, our time sampling
is not frequent enough to allow us to test for lag variability. In the
following, we first derive the light-curves corresponding to both the
accretion disc and hot dust (Section \ref{lcurvesec}). We then use
three different models to infer the lag distribution for the observed
hot dust light-curve (Section \ref{memechosec}), which we find to
converge to a very similar result (Section \ref{revresults}).

\subsection{The observed spectral light-curves} \label{lcurvesec}

\begin{table*}
\caption{\label{acclcurve} 
Near-IR spectral light-curves corrsponding to the accretion disc}
\begin{tabular}{lccccc}
\hline
Observation & HJD & \multicolumn{2}{c}{8700$-$9000~\AA} & \multicolumn{2}{c}{9730$-$9790~\AA} \\
Date & -2,400,000 & original & corrected$^\star$ & original & corrected$^\star$ \\
&& (erg/s/cm$^2$/\AA) & (erg/s/cm$^2$/\AA) & (erg/s/cm$^2$/\AA) & (erg/s/cm$^2$/\AA) \\
\hline
2016 Aug 2  & 57602.76 & (2.87$\pm$0.14)e$-$15 & (2.45$\pm$0.13)e$-$15 & (2.49$\pm$0.07)e$-$15 & (2.12$\pm$0.08)e$-$15 \\
2016 Aug 11 & 57611.74 & (2.42$\pm$0.11)e$-$15 & (2.50$\pm$0.13)e$-$15 & (2.10$\pm$0.08)e$-$15 & (2.16$\pm$0.10)e$-$15 \\
2016 Aug 21 & 57621.76 & (3.96$\pm$0.48)e$-$15 & (2.82$\pm$0.40)e$-$15 & (3.97$\pm$0.45)e$-$15 & (2.83$\pm$0.39)e$-$15 \\
2016 Dec 21 & 57744.08 & (3.24$\pm$0.19)e$-$15 & (2.54$\pm$0.17)e$-$15 & (2.78$\pm$0.10)e$-$15 & (2.18$\pm$0.11)e$-$15 \\
2017 Jan 6  & 57760.10 & (2.44$\pm$0.15)e$-$15 & (2.49$\pm$0.17)e$-$15 & (2.02$\pm$0.11)e$-$15 & (2.06$\pm$0.13)e$-$15 \\
2017 Jan 20 & 57774.09 & (3.19$\pm$0.21)e$-$15 & (2.60$\pm$0.19)e$-$15 & (2.84$\pm$0.11)e$-$15 & (2.31$\pm$0.11)e$-$15 \\
2017 Jan 24 & 57778.08 & (3.31$\pm$0.26)e$-$15 & (2.71$\pm$0.25)e$-$15 & (2.82$\pm$0.16)e$-$15 & (2.31$\pm$0.18)e$-$15 \\
2017 Feb 5  & 57790.09 & (3.96$\pm$0.20)e$-$15 & (2.64$\pm$0.15)e$-$15 & (3.17$\pm$0.09)e$-$15 & (2.11$\pm$0.08)e$-$15 \\
2017 Feb 15 & 57800.11 & (3.27$\pm$0.20)e$-$15 & (2.40$\pm$0.17)e$-$15 & (2.77$\pm$0.11)e$-$15 & (2.03$\pm$0.10)e$-$15 \\
2017 Feb 24 & 57809.10 & (2.32$\pm$0.10)e$-$15 & (2.66$\pm$0.14)e$-$15 & (1.90$\pm$0.05)e$-$15 & (2.18$\pm$0.09)e$-$15 \\
2017 Mar 17 & 57830.01 & (2.28$\pm$0.13)e$-$15 & (2.28$\pm$0.13)e$-$15 & (1.87$\pm$0.08)e$-$15 & (1.87$\pm$0.08)e$-$15 \\
2017 Mar 22 & 57834.99 & (2.81$\pm$0.11)e$-$15 & (2.17$\pm$0.10)e$-$15 & (2.43$\pm$0.07)e$-$15 & (1.88$\pm$0.07)e$-$15 \\
2017 May 9  & 57882.85 & (3.24$\pm$0.17)e$-$15 & (2.89$\pm$0.18)e$-$15 & (2.70$\pm$0.09)e$-$15 & (2.41$\pm$0.11)e$-$15 \\
2017 May 25 & 57898.81 & (2.96$\pm$0.18)e$-$15 & (3.28$\pm$0.23)e$-$15 & (2.40$\pm$0.08)e$-$15 & (2.66$\pm$0.13)e$-$15 \\
2017 Jun 10 & 57914.76 & (3.51$\pm$0.20)e$-$15 & (3.32$\pm$0.22)e$-$15 & (2.77$\pm$0.08)e$-$15 & (2.62$\pm$0.12)e$-$15 \\
2017 Jun 20 & 57924.73 & (3.56$\pm$0.24)e$-$15 & (2.86$\pm$0.22)e$-$15 & (3.08$\pm$0.11)e$-$15 & (2.48$\pm$0.13)e$-$15 \\
2017 Jun 28 & 57932.72 & (4.32$\pm$0.18)e$-$15 & (3.36$\pm$0.18)e$-$15 & (3.50$\pm$0.09)e$-$15 & (2.73$\pm$0.12)e$-$15 \\
2017 Jul 3  & 57937.75 & (3.47$\pm$0.15)e$-$15 & (3.39$\pm$0.20)e$-$15 & (2.91$\pm$0.05)e$-$15 & (2.84$\pm$0.13)e$-$15 \\
\hline
\end{tabular} 

\parbox[]{14cm}{$^\star$ Corrected using the multiplicative factors listed in Table \ref{obslog}, column (13).}

\end{table*}

\begin{table*}
\begin{turn}{90}
\begin{minipage}{220mm}
\vspace*{-0.5cm}
\caption{\label{dustlcurve} 
Near-IR spectral light-curves corresponding to the hot dust}
\begin{tabular}{lccccccc}
\hline
Observation & HJD & \multicolumn{3}{c}{1.55$-$1.60~$\mu$m} & \multicolumn{3}{c}{2.34$-$2.39~$\mu$m} \\
Date & -2,400,000 & original & corrected$^\star$ & corrected & original & corrected$^\star$ & corrected \\
&& (erg/s/cm$^2$/\AA) & (erg/s/cm$^2$/\AA) & (subtracted)$^\dagger$ & (erg/s/cm$^2$/\AA) & (erg/s/cm$^2$/\AA) & (subtracted)$^\dagger$ \\
&&&& (erg/s/cm$^2$/\AA) &&& (erg/s/cm$^2$/\AA) \\
\hline
2016 Aug 2  & 57602.76 & (2.76$\pm$0.04)e$-$15 & (2.35$\pm$0.07)e$-$15 & (1.66$\pm$0.06)e$-$15 & (2.46$\pm$0.04)e$-$15 & (2.10$\pm$0.07)e$-$15 & (1.83$\pm$0.06)e$-$15 \\
2016 Aug 11 & 57611.74 & (2.32$\pm$0.03)e$-$15 & (2.39$\pm$0.07)e$-$15 & (1.70$\pm$0.06)e$-$15 & (2.03$\pm$0.03)e$-$15 & (2.09$\pm$0.07)e$-$15 & (1.82$\pm$0.06)e$-$15 \\
2016 Aug 21 & 57621.76 & (3.81$\pm$0.12)e$-$15 & (2.71$\pm$0.23)e$-$15 & (1.88$\pm$0.17)e$-$15 & (3.32$\pm$0.14)e$-$15 & (2.36$\pm$0.21)e$-$15 & (2.03$\pm$0.19)e$-$15 \\
2016 Dec 21 & 57744.08 & (2.49$\pm$0.04)e$-$15 & (1.96$\pm$0.07)e$-$15 & (1.24$\pm$0.05)e$-$15 & (2.06$\pm$0.04)e$-$15 & (1.61$\pm$0.06)e$-$15 & (1.33$\pm$0.06)e$-$15 \\
2017 Jan 6  & 57760.10 & (1.92$\pm$0.04)e$-$15 & (1.96$\pm$0.07)e$-$15 & (1.28$\pm$0.06)e$-$15 & (1.68$\pm$0.02)e$-$15 & (1.72$\pm$0.06)e$-$15 & (1.45$\pm$0.05)e$-$15 \\
2017 Jan 20 & 57774.09 & (2.41$\pm$0.04)e$-$15 & (1.96$\pm$0.07)e$-$15 & (1.22$\pm$0.05)e$-$15 & (2.03$\pm$0.03)e$-$15 & (1.65$\pm$0.06)e$-$15 & (1.36$\pm$0.05)e$-$15 \\
2017 Jan 24 & 57778.08 & (2.35$\pm$0.05)e$-$15 & (1.93$\pm$0.10)e$-$15 & (1.17$\pm$0.07)e$-$15 & (1.84$\pm$0.03)e$-$15 & (1.51$\pm$0.08)e$-$15 & (1.21$\pm$0.07)e$-$15 \\
2017 Feb 5  & 57790.09 & (2.95$\pm$0.05)e$-$15 & (1.96$\pm$0.06)e$-$15 & (1.26$\pm$0.05)e$-$15 & (2.73$\pm$0.03)e$-$15 & (1.82$\pm$0.05)e$-$15 & (1.54$\pm$0.05)e$-$15 \\
2017 Feb 15 & 57800.11 & (2.52$\pm$0.05)e$-$15 & (1.85$\pm$0.07)e$-$15 & (1.19$\pm$0.05)e$-$15 & (2.23$\pm$0.03)e$-$15 & (1.64$\pm$0.06)e$-$15 & (1.37$\pm$0.05)e$-$15 \\
2017 Feb 24 & 57809.10 & (1.56$\pm$0.03)e$-$15 & (1.80$\pm$0.06)e$-$15 & (1.06$\pm$0.04)e$-$15 & (1.37$\pm$0.02)e$-$15 & (1.57$\pm$0.06)e$-$15 & (1.28$\pm$0.05)e$-$15 \\
2017 Mar 17 & 57830.01 & (1.93$\pm$0.03)e$-$15 & (1.93$\pm$0.07)e$-$15 & (1.32$\pm$0.05)e$-$15 & (1.63$\pm$0.02)e$-$15 & (1.63$\pm$0.05)e$-$15 & (1.40$\pm$0.05)e$-$15 \\
2017 Mar 22 & 57834.99 & (2.60$\pm$0.04)e$-$15 & (2.01$\pm$0.06)e$-$15 & (1.40$\pm$0.05)e$-$15 & (2.30$\pm$0.04)e$-$15 & (1.78$\pm$0.05)e$-$15 & (1.54$\pm$0.05)e$-$15 \\
2017 May 9  & 57882.85 & (2.33$\pm$0.04)e$-$15 & (2.08$\pm$0.07)e$-$15 & (1.26$\pm$0.05)e$-$15 & (1.86$\pm$0.03)e$-$15 & (1.66$\pm$0.06)e$-$15 & (1.34$\pm$0.05)e$-$15 \\
2017 May 25 & 57898.81 & (2.09$\pm$0.03)e$-$15 & (2.32$\pm$0.09)e$-$15 & (1.43$\pm$0.06)e$-$15 & (1.77$\pm$0.03)e$-$15 & (1.96$\pm$0.07)e$-$15 & (1.61$\pm$0.06)e$-$15 \\
2017 Jun 10 & 57914.76 & (2.47$\pm$0.04)e$-$15 & (2.34$\pm$0.10)e$-$15 & (1.44$\pm$0.07)e$-$15 & (1.89$\pm$0.03)e$-$15 & (1.79$\pm$0.07)e$-$15 & (1.44$\pm$0.06)e$-$15 \\
2017 Jun 20 & 57924.73 & (2.83$\pm$0.04)e$-$15 & (2.28$\pm$0.09)e$-$15 & (1.46$\pm$0.07)e$-$15 & (2.29$\pm$0.05)e$-$15 & (1.84$\pm$0.08)e$-$15 & (1.52$\pm$0.07)e$-$15 \\
2017 Jun 28 & 57932.72 & (3.05$\pm$0.03)e$-$15 & (2.37$\pm$0.09)e$-$15 & (1.44$\pm$0.06)e$-$15 & (2.43$\pm$0.06)e$-$15 & (1.89$\pm$0.08)e$-$15 & (1.53$\pm$0.07)e$-$15 \\
2017 Jul 3  & 57937.75 & (2.63$\pm$0.03)e$-$15 & (2.57$\pm$0.11)e$-$15 & (1.64$\pm$0.08)e$-$15 & (2.18$\pm$0.14)e$-$15 & (2.13$\pm$0.16)e$-$15 & (1.77$\pm$0.15)e$-$15 \\
\hline
\end{tabular} 

\parbox[]{18cm}{$^\star$ Corrected using the multiplicative factors listed in Table \ref{obslog}, column (13).}

\parbox[]{18cm}{$^\dagger$ Corrected using the multiplicative factors listed in Table \ref{obslog}, column (13), and with the accretion disc spectrum subtracted.}

\end{minipage}
\end{turn}
\end{table*}

We have assembled the light-curves of both the accretion disc (Table
\ref{acclcurve}) and the hot dust component (Table \ref{dustlcurve})
from fluxes measured in the near-IR spectrum. Ideally, we would like
to measure the accretion disc flux at the shortest wavelengths, where
it is least contaminated by the hot dust component. However, the
$S/N$~ratio of the near-IR spectra decreases significantly towards
short wavelengths and is relatively low in particular in the region
$<0.8~\mu$m ($S/N \sim 5$). Therefore, we have chosen to measure the
accretion disc flux in the 60~\AA~wide rest-frame wavelength region of
\mbox{$\lambda=9730-9790$~\AA}, which lies between the two broad
hydrogen emission lines Pa$\epsilon$ and Pa$\delta$ and is known to be
line-free, and also in the wider (300~\AA) rest-frame wavelength
region of \mbox{$\lambda=8700-9000$~\AA}, which does not appear to be
contaminated by significant emission-line flux in NGC~5548 (see
Fig.~\ref{irtfspec}). We have measured the average flux in the
selected wavelength ranges and have calculated error bars in the usual
way as the standard error on the flux average. We note that these
errors are only a few per cent and so on average much smaller than the
applied photometric correction factors.

We have measured the flux of the hot dust in the near-IR spectrum {\it
  after} the subtraction of the accretion disc component. This is
important not only because the accretion disc flux dominates almost
the entire $J$ band, but also because the accretion disc contribution,
although it decreases with increasing wavelength, remains
non-negligible even at the long-wavelength end of the $K$ band, where
the hot dust spectrum peaks (e.g. $\sim 15\%$ of the total flux at
$2.4~\mu$m in the observation from 2017 March 17; see Table
\ref{dustlcurve}). Therefore, we have chosen to measure the flux of
the hot dust in two line-free, 500~\AA~wide rest-frame wavelength
regions, namely, \mbox{$\lambda=2.34-2.39~\mu$m}, which is at the red
end of the $K$ band but close to the peak of the blackbody spectrum,
and \mbox{$\lambda=1.55-1.60~\mu$m}, which is approximately in the
middle of the $H$ band.

During this campaign, NGC~5548 was in a relatively low emission state
(on average a factor of $\sim 2$ lower than during the 2014 {\it HST}
campaign), but was highly variable (Fig.~\ref{lcurves}). The continuum
fluxes at the shortest wavelengths corresponding to the accretion disc
had maximum variations of \mbox{$\sim 50 - 60\%$} and the light-curves
show several clearly isolated peaks and dips. This high variability of
the observed light-curves is a prerequisite for recovering lag times
from the data.

\subsection{The \memecho~and \mcmcrev~formalisms} \label{memechosec}

\memecho~\citep{Horne91, Horne94} is a formalism based on the
maximum-entropy method and infers the lag distribution given a driver
and a reprocessor light-curve. In short, \memecho~fits a linearised echo model:

\begin{eqnarray}
F_{\rm uv}(t) & = & \overline{F}_{\rm uv} + \Delta F_{\rm uv}(t) \nonumber\\
F_{\nu}(t) & = & \overline{F}_{\nu} + \Delta F_{\nu}(t) \\
F_{\nu}(t) & = & \overline{F}_{\nu} + \int{\Psi_\nu(\tau) \Delta F_{\rm uv}(t- \tau) d\tau} \nonumber
\end{eqnarray}

\noindent
where both the observed driver light-curve, $F_{\rm uv}(t)$, and the
observed reprocessor light-curve, $F_{\nu}(t)$, are assumed to be
composed of a constant and a variable component. The total flux of the
reprocessor is then modelled as the sum of the constant (mean)
reprocessor flux, $\overline{F}_\nu$, and the convolution of the
response function, $\Psi_\nu(\tau)$, with the variable part of the
driver light-curve, $\Delta F_{\rm uv}(t)$. \memecho~recovers from the
data the union of the three positive functions, $\Psi_\nu(\tau)$,
$F_{\rm uv}(t)$ and $\overline{F}_\nu$. The total entropy of the image
is the sum of the entropies of these three subimages and of the many
different images that satisfy the data constraint, the image with
maximum entropy is selected.

\mcmcrev~\citep[similar to CREAM;][]{Starkey16} uses the MCMC
technique to sample posterior parameter distributions for a model
response function given the light-curves for the driver and the
reprocessor. We have modelled the response function with a log-normal
distribution and with a temperature-radius relationship for a standard
accretion disc seen face-on (\mbox{$T \propto R^{-3/4}$}). The latter
is motivated by recent models of the dusty torus as a radiatively
accelerated outflow launched from the outer regions of the accretion
disc \citep{Czerny11, Czerny17}. In effect, the temperature-radius
relationship translates into a wavelength-lag relationship.

\subsection{Reverberation results} \label{revresults}

\begin{figure*}
\centerline{
\includegraphics[angle=-90,scale=0.5]{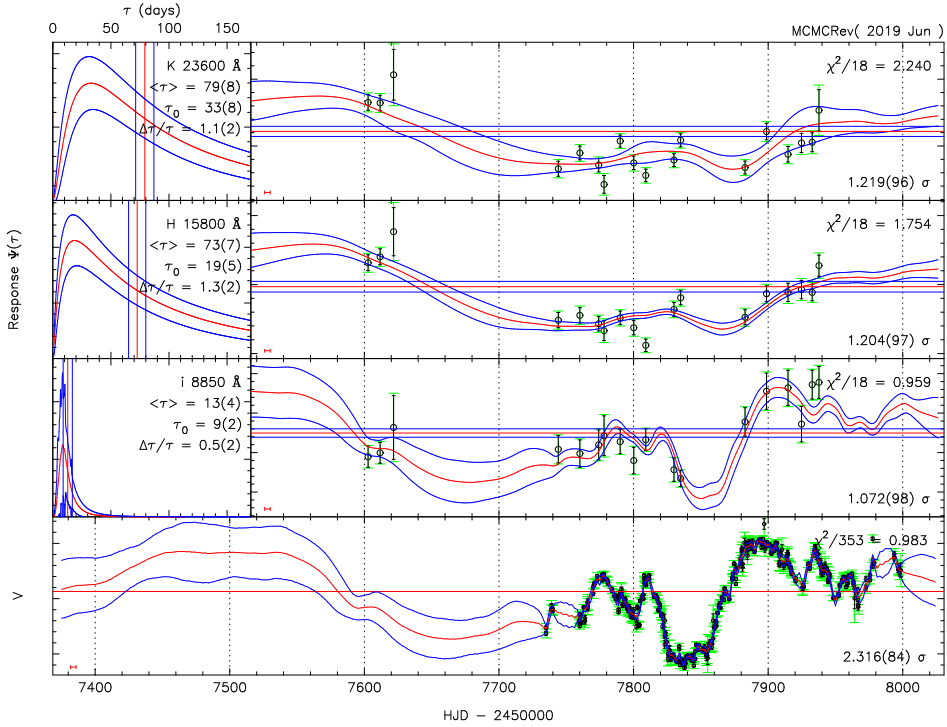}
}
\caption{\label{mcmcrevlognorm} \mcmcrev~results for a driving
  light-curve fitted to the $V$-band photometry (bottom panel) and
  echo light-curves for log-normal delay distributions fitted to the
  $K$-band (top panel), $H$-band (second top panel) and
  $8700-9000$~\AA (third top panel) spectroscopic light-curves. Error
  envelopes for the light-curves (right-hand panels), background
  levels and delay maps (left-hand panels), shown as red curves
  bracketted by blue curves, are the mean and rms of the MCMC
  samples. The vertical lines in the delay maps indicate the mean and
  its error, which are $\langle \tau \rangle = 79\pm8$, $73\pm7$ and
  $13\pm4$ for the $K$-band, $H$-band and $8700-9000$~\AA,
  respectively.}
\end{figure*}

\begin{figure}
\centerline{
\includegraphics[scale=0.46]{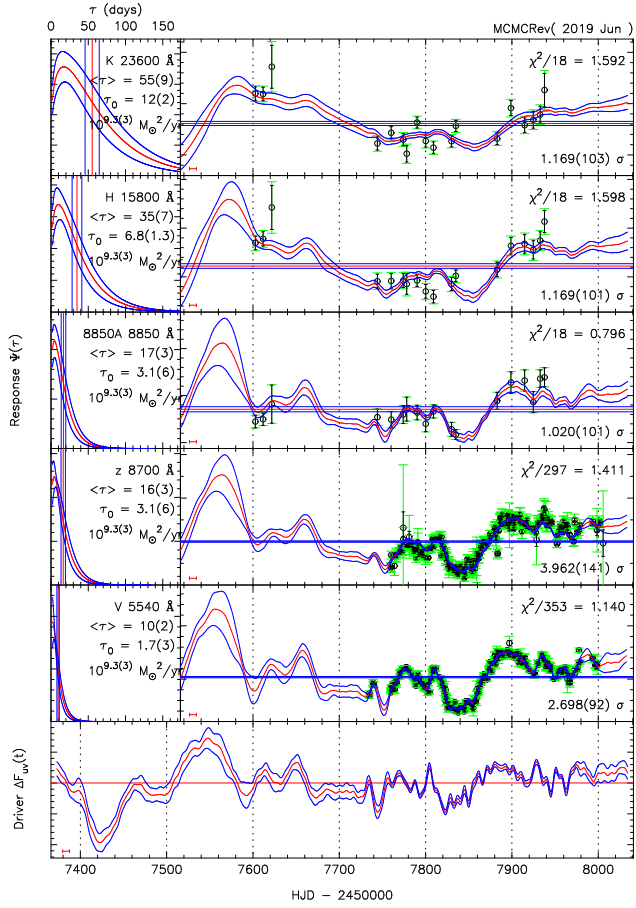}
}
\caption{\label{mcmcrevdisc} \mcmcrev~results assuming the response
  function of a standard accretion disc seen face-on. Symbols are as
  in Fig. \ref{mcmcrevlognorm}. The results are summarised in Table
  \ref{mcmcrevacc}.}
\end{figure}

We have run \memecho~with both the $8700-9000$~\AA~spectroscopic
light-curve and the $V$ band photometric light-curve as the
driver. The delay distributions for both the $H$ and $K$ dust
light-curves have a clear peak, which is at $\sim 40-45$~days and at
$\sim 70-80$~days using the spectroscopic and photometric light-curves
as the driver, respectively. The width of the \memecho~delay map is
much narrower in the former case, most likely due to the spectroscopic
light-curve being too smooth.

For an MCMC fit, posteriors on the model parameters can be interpreted
as measurements with error bars. In this case, the centroid of the
response function gives the average response-weighted dust radius,
which is defined as:

\begin{equation}
\langle R_{\rm d,rev} \rangle = c \langle \tau \rangle = \frac{\int c \tau \Psi(\tau) d\tau}{\int \Psi(\tau) d\tau}
\end{equation}

\noindent
Using the $V$ band light-curve as the driver and modelling the
response function with a log-normal distribution, we get similar mean
values for the dust delay map in the $H$ and $K$ band of
\mbox{$\langle R_{\rm d,rev} \rangle = 73 \pm 7$~light-days} and
\mbox{$79 \pm 8$~light-days}, respectively
(Fig. \ref{mcmcrevlognorm}). But sparse time sampling and relatively
large error bars on the infrared light-curve data limit the
information that can be obtained on the delay distribution. In this
situation, the mean delay can be constrained fairly well, but not the
detailed shape. With this caveat, the log-normal delay maps for the
$H$ and $K$ bands peak at $\sim 20$~and 35~days, respectively. We note
also that evidence for a lack of prompt response in $H$ and $K$ occurs
just after the $V$ band flux has dropped sharply between days 7810 and
7825. In response to this the $8700-9000$~\AA~flux has dropped by day
7830, but the $H$ and $K$ fluxes remain high at days 7810 and 7835.

Assuming instead for the response function the temperature-radius
relationship of a standard accretion disc ($\tau \propto
\lambda^{4/3}$) does not change these values significantly
(Fig. \ref{mcmcrevdisc}). We get average values of \mbox{$\langle
  R_{\rm d,rev} \rangle = 35 \pm 7$~light-days} and \mbox{$55 \pm
  9$~light-days} for the $H$ and $K$ band, respectively (Table
\ref{mcmcrevacc}). However, the results show consistency with the
accretion disc temperature-radius relationship for both the average
and peak values. Moreover, the delay distribution is strongly skewed
towards low values with the peak at \mbox{$\tau_0 = 7 \pm
  1$~light-days} and \mbox{$12 \pm 2$~light-days} for the $H$ and $K$
band, respectively. We discuss this further in Section \ref{geometry}.

\begin{table}
\caption{\label{mcmcrevacc} 
Results from \mcmcrev~assuming the response function of a standard accretion disc}
\begin{tabular}{llccc}
\hline
Wavelength & band & $\tau_0$ & $\langle \tau \rangle$ & $\chi_\nu^2$ \\
(\AA) && (lt-days) & (lt-days) \\
\hline
23600 & $K$ (spectrum) & 12$\pm$2 & 55$\pm$9 & 1.59 \\
15800 & $H$ (spectrum) &  7$\pm$1 & 35$\pm$7 & 1.60 \\
8850 & $z$ (spectrum) & 3$\pm$6 & 17$\pm$3 & 0.80 \\
8700 & $z_s$ (photometry) & 3$\pm$6 & 16$\pm$3 & 1.41 \\
5448 & $V$ (photometry) & 2$\pm$3 & 10$\pm$2 & 1.14 \\
\hline
\end{tabular} 
\end{table}

\section{The torus in NGC~5548} \label{torus}

There are three ways to measure or estimate the radius of the hot dust
in AGN: (i) through the time delay in the response of the dust
emission to the central heating source; (ii) assuming radiative
equilibrium for known temperature and grain properties, and (iii)
measuring the distance of the dust emission by geometrical means,
e.g. with near-IR interferometry or high spatial-resolution
imaging. In general, it is found that interferometric dust radii are a
factor of $\sim 2$ larger than reverberation dust radii \citep{Kish09,
  Kish11, Kosh14}. Since the former are effectively
luminosity-weighted dust radii, this result can be understood if one
assumes that response-weighted radii are dominated by the fastest
variations, which are expected from the smallest radii, whereas
luminosity-weighted radii are dominated by the emission from grains
with the highest emissivity, i.e. the hottest and largest grains. The
highest temperature of the largest grains will, in any case, be lower
than the highest temperature of the smallest grains, due to their
lower heat capacity, and so their emission will come from further
out. Alternatively, \citet{Kaw10, Kaw11} ascribe the small
reverberation radii to a dust geometry that is concave (or
bowl-shaped) rather than spherical due to the anisotropy of the
accretion disc emission; since the UV/optical accretion disc
luminosity is reduced for larger viewing angles
\citep[e.g.][]{Netzer87, Hubeny00}, so is the dust sublimation radius
(on average by a factor of $\sim 2-3$). The interferometric radii
would then be dominated by the region containing the bulk of the dust,
which in this geometry is located further out. Comparisons between
reverberation dust radii and luminosity-weighted dust radii derived
from SEDs have yielded so far ambiguous results \citep{Mor12, L14,
  Kosh14}, most likely due to the unknown grain properties and a
poorly constrained dust temperature and irradiating luminosity. In
this study, we combine for the first time the simultaneous measurement
of a reverberation radius and of a well-constrained luminosity-based
radius and show that it can constrain the grain size (Section
\ref{location}), the chemical composition (Section \ref{dynamics}) and
the origin (Section \ref{geometry}) of the dust.

\subsection{The location} \label{location}

Our response-weighted radius of $\sim 70$~light-days (Section
\ref{revresults}) is only $\sim 15\%$ larger than the average
luminosity-weighted radius for dust of large grains, but much smaller
than the average luminosity-weighted radius for dust of small grains
(by a factor $\ga 6$ and $\ga 8$ for carbon and silicate dust,
respectively; Section \ref{lumradius}). Taking into account the
uncertainty of the irradiating luminosity in the calculation of the
luminosity-weighted radius gives that in the blackbody case the
reverberation radius is larger than the average luminosity-based
radius by a factor of $\sim 3$ and in the case of small-grain carbon
and silicate dust smaller by a factor $\ga 2$ and $\ga 3$,
respectively. Clearly, the smallest discrepancy between the response-
and luminosity-weighted radius is for dust of large grains and for an
irradiating luminosity as estimated.

Given that we have almost certainly overestimated the UV/optical
accretion disc luminosity (see Section \ref{accspectrum}), the
similarity between the reverberation and luminosity-based dust radius
in the case of a blackbody emissivity law is intrigueing. A reason for
this could be that the dust is also heated by X-rays. For small
grains, the absorption efficiency drops sharply for frequencies larger
than the UV, but for large grains it continues to be
high. \citet{Meh15a} have presented a well-sampled, multiwavelength
SED of NGC~5548, which extends to hard X-ray frequencies and was
corrected for host galaxy emission but included the `constant red
component' (see their Fig.~6, right panel). Scaling this SED to our
$8700-9000$~\AA~continuum flux gives an irradiating luminosity $\sim
40-70\%$ higher than our (over)estimated accretion disc luminosity,
which translates into an increase in dust radius of $\sim 20 - 30\%$
and so an average radius for the blackbody case of $\sim
70-80$~light-days. This plausible match with the reverberation radius
indicates that X-rays are important in heating the dust and constrains
the grain size further to $a \ga 2~\mu$m for both graphite and
silicates \citep{Draine16, Draine17}.

It was previously proposed that the dust composition in the
circumnuclear regions of AGN is dominated by large grains (of a few
$\mu$m). \citet{Laor93} found that small grains in an optically thin
dust configuration are likely to be destroyed on short timescales in
an AGN environment, whereas very large grains are not. Large grains in
AGN were also postulated by \citet{Mai01b} to explain the observed
lack of prominent silicate absorption features in the mid-IR spectra
of Seyfert 2s and the lack of a strong 2175~\AA~(carbon) feature in
the UV spectra of reddened Seyfert 1s, since they make the extinction
curve flatter and featureless. Furthermore, the high-density
environment at the centre of an AGN is expected to be a natural
catalyst for the formation of large grains by coagulation. On the
other hand, almost certainly not all dust is concentrated in large
grains as shown by IR polarisation studies, which are sensitive to the
smallest dust grains and give constraints of $a \la 0.1~\mu$m
\citep[e.g.][]{Lopez13, Lopez17}.

\subsection{The dynamics} \label{dynamics}

\begin{figure}
\centerline{
\includegraphics[scale=0.42]{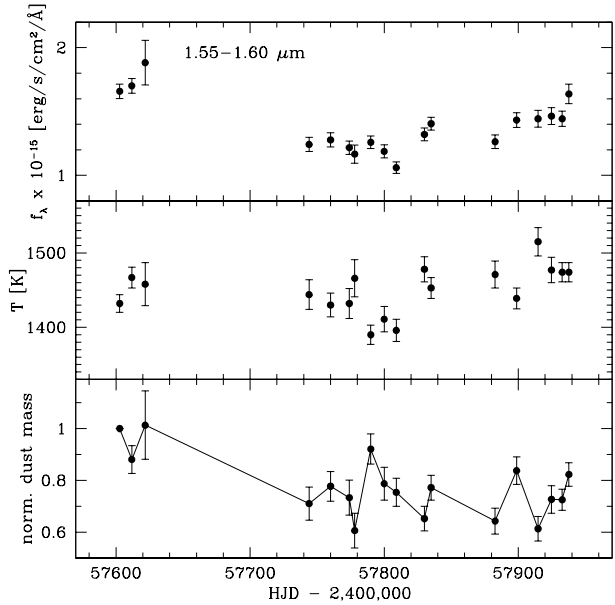}
}
\caption{\label{tempcurveplot} Temporal behaviour of the dust flux in
  the $H$ band (top panel), dust temperature (for the blackbody case;
  middle panel) and normalised dust mass (relative to the start of the
  campaign and connected in time for clarity; bottom panel). We plot
  $1\sigma$ errors.}
\end{figure}

\begin{figure}
\centerline{
\includegraphics[scale=0.45]{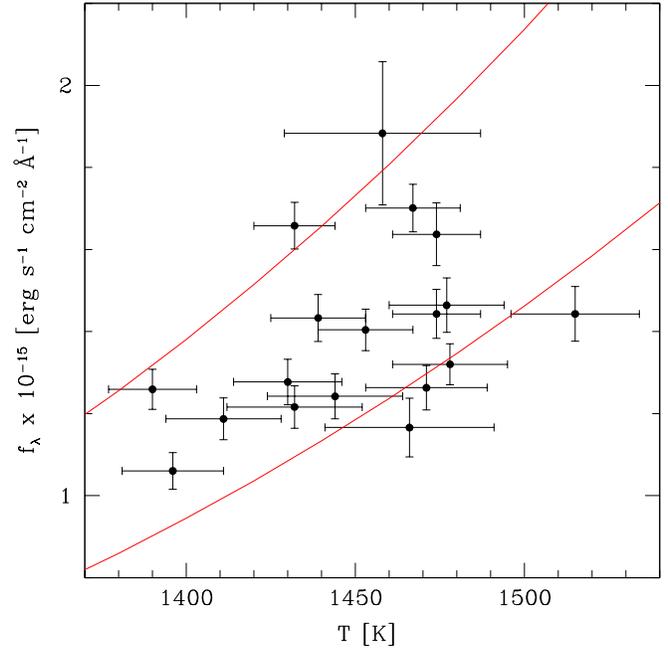}
}
\caption{\label{tempfdust} Flux in the wavelength region of
  $1.55-1.60~\mu$m versus the temperature (for the blackbody case). We
  plot $1\sigma$ errors. The red curves show the expected relationship
  if the flux increase is solely due to a temperature increase. The
  spread in dust mass necessary to explain the flux difference between
  the two red curves is a factor of $\sim 1.5$.}
\end{figure}

It is usually assumed that the inner edge of the torus is set by
thermal sublimation processes, which implies that dust can exist as
far in as is physically possible without being destroyed via direct
heating by the central irradiating source, and that the dust radius
changes mainly in response to the AGN luminosity \citep[the `receding
  torus model';][]{Law91}. We can now test this assumption with our
measured dust temperatures and their variation with time, since, as we
have shown in Section \ref{accspectrum}, the most variable dust
component also dominates the emission. If the dust is close to
sublimation, the observed temperature can constrain the chemical
composition of the dust, since the expected sublimation temperatures
are very different for carbon and silicate dust \citep[$T_{\rm sub}
  \sim 1800 - 2000$~K and $\sim 1300 - 1500$~K,
  respectively;][]{Sal77, Lod03}. If the hot dust is much cooler than
its sublimation temperature, this implies that the dust-free inner
region in AGN is enlarged and that the inner edge of the torus is set
by a different process and so maybe independent of the irradiating
luminosity. In this respect we note that if the central luminosity is
powerful enough to heat the dust to sublimation, the dust temperature
will generally be close to sublimation and there will not be an
enlarged dust-free inner region ``evacuated'' by a previous high-state
in luminosity stable on time-scales longer than grain reformation
times (i.e. a few months). In addition, since $L_{\rm uv} \propto T^4$
(eq. \ref{Stefan-Boltz}), only a strong increase in luminosity (by
factors of a few) and lasting for several months (i.e. much longer
than typical grain reformation times) could have produced the current
radius at which the dust temperature is now well below
sublimation. This required variability behaviour is not usually
observed for radio-quiet AGN. In particular for NGC~5548, large UV
flux variations occur only over shorter periods of weeks
\citep[e.g.][]{Storm1}.

We have found a temperature range of \mbox{$T \sim 1390 - 1515$~K} for
the blackbody case\footnote{We will consider here and in the following
  only the blackbody case, since we have shown in (Section
  \ref{location}) that the nuclear dust in NGC~5548 is composed mainly
  of large grains.} (Section \ref{lumradius}), which is consistent
with silicates {\it at} sublimation and is $\sim 300 - 500$~K below
the sublimation temperatures of carbon dust. Since the growth and
survival of dust particles requires low enough grain temperatures
(well below the condensation limit), this result indicates that the
hot dust consists mainly of carbon, with silicates having sublimated.
This conclusion is also supported by previous near-IR spectra of
NGC~5548, where blackbody dust temperatures as high as \mbox{$T \sim
  1700$~K} were found \citep{L15b}. The fact that we have measured a
dust reverberation signal in this campaign also points to the dust
being well below sublimation so that further heating is possible
without destroying the dust. Both temperature and mass changes (e.g.,
due to grain destruction) can introduce variations in the dust
flux. However, assuming instantanous heating, the change in dust flux
due to a change in temperature according to eq. \ref{Stefan-Boltz} is
expected to dominate the variations since grain reformation times are
relatively large (of the order of a few months). Then, if the dust was
too close to sublimation, only a change in dust mass could introduce a
significant change in dust flux. But such a flux change will not
resemble a reverberation signal since it will not be the delayed and
smoothed version of the driver light-curve. Since sublimation
timescales are short and sublimation is a runaway process once the
grain starts to evaporate, an increase in irradiating luminosity will
quickly {\it decrease} the dust flux and this will stay low and
unchanged until grains have reformed and the flux can increase
again. But grain reformation not only takes a relatively long time, it
also requires the irradiating luminosity to stay continuously low for
that (prolonged) period.

In Fig.~\ref{tempcurveplot} we show the temporal behaviour of the dust
flux and temperature. The two are very similar, i.e. the change in
dust flux during our campaign is mainly due to a change in
temperature, and the temperature variations are consistent with direct
heating by the accretion disc. We get a $1\sigma$ dispersion around
the mean \mbox{$8700 - 9000$~\AA}~continuum flux of $\sim 13\%$
\mbox{($\langle f_\lambda \rangle = (2.74 \pm 0.38) \times
  10^{-15}$~erg~s$^{-1}$~cm$^{-2}$~\AA$^{-1}$)}, based on which we
expect a dispersion around the mean temperature of $\sim 3\%$. The
observed dispersion is $\sim 2\%$ (\mbox{$\langle T \rangle = 1450 \pm
  32$~K}). Based on the maximum variation in the $8700 -
9000$~\AA~continuum flux of $\sim 56\%$, we expect a difference
between the minimum and maximum temperature of $\sim 12\%$, which
translates to $\sim 170$~K, similar to the observed value of $\sim
130$~K.

If it is not sublimation, then what sets the location of the inner
edge of the torus? Or, put differently, why is the hot dust located
much further out and so the dust-free inner region enlarged? There are
two possibilities; either there was initially carbon dust in the inner
regions permitted by sublimation but it has since been destroyed by
localized processes or this inner region has always been
dust-free. The latter possibility is unlikely if gas is to be expected
in this region, since where there is gas there usually is
dust. Furthermore, the gas densities in the inner region are expected
to be higher than further out, which favours grain formation and
growth. Therefore, most likely, dust destruction has taken place. If
so, we would expect this process, which should be observable as a
change in dust mass, to also affect the dusty region to some degree,
since its efficiency would drop only gradually with radius.

In Fig.~\ref{tempfdust}, we show the dust flux versus the dust
temperature. Also plotted are heating tracks that show the expected
flux increase due to a temperature increase alone (i.e. for a constant
dust mass). It is clear that the data cannot be fit by a single such
heating track, i.e. the spread in flux is larger than what is expected
from a change in temperature alone, and so a changing dust mass has to
be assumed additionally. In Fig.~\ref{tempcurveplot} (bottom panel) we
show the normalised dust mass relative to the start of the campaign
and connected in time. As expected for a grain destruction and
reformation scenario, the dust mass is the highest at the beginning of
the campaign. This is followed by the largest mass loss (by $\sim
40\%$, significant at the $6\sigma$ level), from which the dust mass
recovers close to its initial value after $\sim 5.5$~months, followed
by a renewed mass loss (again by $\sim 40\%$, significant at the
$7\sigma$ level), from which the dust starts to recover only at the
end of our campaign, i.e. again after $\sim 5$~months. Our estimate of
a grain reformation time of $\sim 5-6$~months is within the usual
theoretical range and of the same order as that estimated for
Fairall~9 by \citet{Barv92} (a few months) and for NGC~4151 by
\citet{Kosh09} (about a year). In this respect, we note that if the
hot dust was close to its sublimation temperature, we would also
expect a decrease in dust mass due to grain evaporation, but this
would not necessarily be observable as a decrease in near-IR flux. For
a large distribution in grain size, sublimation will affect mainly the
grains with the smallest sizes, since they have the smallest heat
capacity, but these grains do not dominate the emission. If the grain
size distribution is relatively narrow, as we find in our case, all
grains will be similarly affected by sublimation. But, since
sublimation timescales are short \citep{Baskin18}, this loss of dust
mass loss will not only be drastic but sudden. Finally, we note that
our observation of a dust mass change that is not accompanied by large
variations in dust temperature further indicates that the enlarged
dust-free inner region was not ``evacuated'' by a previous high-state
in luminosity and is now being replenished, but that it is rather kept
dust-free by on-going in-situ processes. We will return to this point
in Section \ref{geometry}.

Temperature estimates in photometric dust reverberation campaigns are
rare since they require simultaneous observations in several filters
and in any case have a high uncertainty \citep{Clavel89,
  Glass04}. Temperature variations in such campaigns have only
recently been studied. \citet{Schnuelle13, Schnuelle15} performed a
5-month photometric monitoring campaign on NGC~4151 with GROND
covering the $J$, $H$ and $K$ bands. They also found large blackbody
temperature variations (of $\sim 200$~K) and changes in covering
factor (both a decrease and increase).

Our result that the hot dust is not close to sublimation predicts that
the location of the inner edge of the torus is largely independent of
the irradiating luminosity and so rather a ``torus wall''. This is
confirmed for NGC~4151 by the observed variability of both the
interferometric and reverberation dust radius. \citet{Pott10}
presented a comprehensive analysis of all then available near-IR
interferometric observations and concluded that the hot dust radius is
largely luminosity-invariant. \cite{Kosh09} measured the reverberation
radius for eight separate epochs over a six-year period and found it
to vary largely independent of luminosity, making dust destruction and
reformation a necessary part in the explanation of the
observations. The length of our near-IR monitoring campaign covers in
principle the mean reverberation lag several times, but,
unfortunately, our time sampling is not frequent enough to allow us to
test for lag variability.

Finally, we note that, as discussed by \citet{Baskin18}, the dust
sublimation temperature also depends weakly on the gas number
density. The sublimation temperatures assumed here for graphite and
silicate are relevant for the high-density gas expected in the torus
(of $n_e \sim 10^7-10^9$~cm$^{-3}$). If we assume that graphite was at
sublimation at $T\sim 1450$~K, it would imply a density of only
\mbox{$n_e \sim 10^4$~cm$^{-3}$} (see their eq. (32)). But such
rarified gas at a distance of $\sim 70$~light-days would be very hot
if it was exposed to the total bolometric luminosity and the dust
grains would be destroyed by thermal sputtering. An interpretation of
our results in terms of silicate dust being close to sublimation would
imply that it is carbon-poor. However, amorphous carbon as the main
constituent of the hot dust is required by dust-driven wind models
since it dominates the overall dust opacity and so provides the
driving force of the wind. We will expand on this point in the next
Section \ref{geometry}. In any case, it lies at hand to associate the
enlarged dust-free inner region with the coronal line region
\citep{Pier95} and we plan to explore this possibility in the future.

\subsection{The geometry} \label{geometry}

\begin{figure}
\centerline{
\includegraphics[scale=0.45]{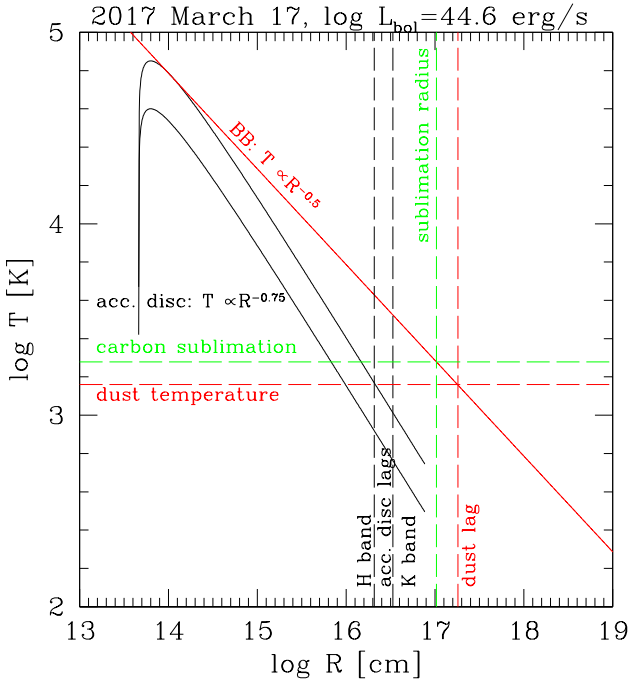}
}
\caption{\label{tempradius} The temperature-radius relationships for
  the accretion disc (black curves) and hot dust (red curve) for the
  observation from 2017 March 17. The red horizontal and vertical
  dashed lines indicate the observed dust temperature ($T \sim
  1450$~K) and reverberation radius ($R_{\rm d,rev} \sim
  70$~light-days), respectively. The green horizontal and vertical
  dashed lines indicate the expected carbon sublimation temperature
  ($T \sim 1900$~K) and radius ($R_{\rm d,sub} \sim 40$~light-days),
  respectively. The black vertical dashed lines indicate the peak lag
  times in the H and K bands assuming for the response function the
  temperature-radius relationship of a standard accretion disc ($\sim
  7$ and 12~light-days, respectively).}
\end{figure}

In recent years, an understanding of the tight connection between the
extent of the BLR and the innermost part of the dusty torus has
started to emerge \citep{Czerny11, Goad12, L14}. This connection is
made in particular within models for radiatively accelerated dusty
outflows launched from the outer regions of the accretion disc
\citep[e.g.][]{Elitzur06, Czerny16, Czerny17}. In this case, the large
opacity of carbon dust leads to an inflated disc structure of which
the (cold) back side naturally forms a hot dusty torus
\citep{Baskin18}. But this hot dust, will not necessarily be
externally illuminated, since it lies largely in the plane of the
accretion disc, and so it will not reverberate in the classical
sense. In addition to dust located within a toroidal structure and to
that condensing out in the accretion disc material, a third component
could be present in the polar rather than the equatorial region if the
polar dust observed at mid-IR wavelengths \citep{L10b, Hoenig13,
  Fuller19} extends close into the nucleus. Our results can constrain
which of these three hot dust components is observed.

We start by using our simultaneous measurement of the temperature and
reverberation radius together with its emissivity law to ``calibrate''
the temperature-radius relationship for the hot dust. We show this
relationship in Fig.~\ref{tempradius} (red solid line) for the
observation from 2017 March 17. This temperature-radius relationship,
which is that of a blackbody and so of the form $T \propto R^{-0.5}$,
is that of dust exposed at all times to the total bolometric
luminosity. The bolometric luminosty for this observation is
\mbox{$\log L_{\rm bol} \sim 44.6$~erg~s$^{-1}$}, if we scale the
multiwavelength SED of NGC~5548 presented by \citet{Meh15a} to our
$8700-9000$~\AA~continuum flux. We then compare the temperature-radius
relationship of the hot dust to that for the accretion disc, which is
of the standard form $T \propto R^{-0.75}$, and so steeper than that
for the hot dust (black solid line). We have considered both the
accretion disc luminosity that approximates the observed continuum
flux as well as the lowest possible luminosity, with the latter lower
than the former by a factor of $\sim 10$, due to the presence of the
`constant red component' (Section \ref{accspectrum}). From these
curves, one can immediately see that dust is expected to condense out
in the accretion disc matrial at radii much smaller than those
permitted by the total bolometric luminosity. Then, if this dust was
to be raised off the accretion disc to considerable heights, e.g., by
a dusty outflow, it would immediately be destroyed, as discussed in
\citet{Czerny17}. But, if this dust exists, it would emit thermal
near-IR radiation and we should be able to observe it. If it is carbon
dust, we expect it in our case at a radius of \mbox{$\sim
  3-5$~light-days} if it is close to sublimation ($T \sim 1900$~K) and
at a radius of \mbox{$\sim 4-9$~light-days} if it is at our observed
temperature ($T \sim 1450$~K). Assuming for the response function the
temperature-radius relationship of a standard accretion disc, we
obtained in Section \ref{revresults} average delay times much larger
than this, but peak values of $\sim 7\pm1$ and $12\pm2$~light-days for
the $H$ and $K$ band, respectively (vertical black dashed lines in
Fig.~\ref{tempradius}). Therefore, within the errors, this
reverberation signal is consistent with dust in the accretion
disc. But if so, this implies relatively large accretion disc outer
radii of at least \mbox{$R_{\rm out} \sim 1700-3000\,r_{\rm g}$}.

Our observation of a changing dust mass implies that on short
time-scales (of about a year) the mass loss rate is not balanced by
the mass inflow rate, i.e. the inner edge of the torus is {\it not} in
a steady-state. This is expected in wind scenarios for the torus
\citep{Koenigl94, Elitzur06, Keating12}. But even in the hydrostatic
picture, radiation pressure is expected to dominate over gravity at
the inner edge and so clouds in this (small) region will not be able
to maintain bound orbits \citep{Kro86}. The observed loss in dust mass
could also be due to dust destruction (of the largest grains) by
in-situ processes. Possible dust destruction processes that could
happen locally, e.g., in fast shocks, are grain sputtering caused by
collisions with high-energy ($>10$~eV) atoms or ions and grain
fragmentation (or even vaporization) through grain-grain collisions,
which are expected to almost completely remove large grains ($a \ge
0.1~\mu$m) \citep{Jones04}. Fast shock waves are naturally expected in
the outer accretion disc through the dissipation of turbulence,
dynamical instabilities or spiral shocks. In the case of NGC~5548, the
fast ($\sim 1000-5000$~km~s$^{-1}$) and long-lived outflow from
locations just outside the BLR detected by \citet{Kaastra14} during
the 2013 {\it HST/XMM-Newton} monitoring campaign could also be a
source for fast shocks.

Finally, if there was hot polar dust present on subarcsecond scales
\citep{Hoenig13}, in addition to the dusty torus, we would have a
situation similar to that described in \citet{Kaw10} (see their
Fig.~2) and \citet{Baskin18} (see their Fig.~13); due to the
anisotropy of the irradiating (accretion) disc emission ($L_{\rm bol}
\propto \cos \theta$, with $\theta$ the angle between the accretion
disc rotation axis and the location of the dust), the dust in the
polar regions will be located further out than dust in the equatorial
plane if both have the same temperature. How much further out the
polar dust will be depends on the scale height of the torus, i.e. on
its average $\theta$. If we assume for this the usual value of $\theta
\sim 60-70^\circ$, the polar dust would be $\sim 50\%$ further away
than the torus dust. But this radius difference will increase fast as
the average $\theta$ of the torus increases further, as would be
expected, e.g. for the rather flat dusty structure of
\citet{Baskin18}. Clearly, if the polar and torus hot dust have the
same temperature, the polar dust is not expected to dominate the
variations. If we assume instead that the polar dust is hotter than
the hot torus dust, it will have to dominate both the emission and the
variations, since in our campaign we observe the variable (rms)
spectrum and the mean spectrum to have the same spectral shape at
wavelengths $\ga 1~\mu$m (Section \ref{accspectrum}). This would then
imply that the hot dust in the torus is at much lower temperatures ($T
<< 1000$~K) in order for it to not dominate the variations (if not the
emission). The emission of such dust peaks at wavelengths well outside
our near-IR spectral range ($\lambda >> 3~\mu$m) and so we would not
have observed it. But, in this case, it would be difficult to explain
why there is hot polar dust but not hot torus dust (at $T \sim
1450$~K).

\section{Summary and conclusions}

We have conducted the first spectroscopic dust reverberation programme
on the well-known AGN NGC~5548. Spectroscopy, unlike photometry, can
simultaneously determine several dust properties, such as its flux,
temperature and covering factor. In particular, it can measure both
response-weighted and luminosity-weighted dust radii. Our main results
can be summarised as follows.

\vspace*{0.2cm}

(i) We present the first near-IR variable (rms) spectrum for an AGN
and find that it shows the usual spectral inflection at $\sim 1~\mu$m,
indicating that both the accretion disc and the hot dust contribute to
the variability. Most importantly, we find that the spectral shape at
wavelengths $\ga 1~\mu$m is similar in the variable (rms) and mean
spectrum, which means that the same dust component dominates both the
emission and the variations. Based on this, we constrain the hot dust
properties.

(ii) We measure the response time in the near-IR using several
techniques and find them to converge to a value of $\sim
70$~light-days. This is at the high end of the range of $\sim 40 -
80$~light-days measured previously by photometric reverberation
campaigns. Assuming thermal equilibrium for optically thin dust, we
derive the luminosity-based dust radius using our measurement of the
temperature and an estimate of the irradiating luminosity, both
obtained from the near-IR spectrum. We find that the two dust radii
are in excellent agreement if we assume a wavelength-independent dust
emissivity-law, i.e. a blackbody, which is appropriate for grains of
relatively large sizes (of a few $\mu$m).

(iii) We measure an average dust temperature of \mbox{$T \sim
  1450$~K}, which indicates that silicates have sublimated
\mbox{($T_{\rm sub} \sim 1300-1500$~K)} and graphite dominates the
chemical composition. However, the measured temperature is several
100~K below the sublimation temperature of astrophysical carbon
\mbox{($T_{\rm sub} \sim 1800-2000$~K)} and we observe temperature and
flux variations for the hot dust consistent with it experiencing
heating and cooling by the central irradiating source. Since the
observations also require the dust mass to change, which we find is
destroyed by up to a maximum of $\sim 40\%$ and reforms on time scales
of $\sim 5-6$~months during our campaign, this would imply that the
inner dust-free region is enlarged by in-situ dust destruction
proceses. This then means that the inner radius of the dusty torus is
not set by sublimation and is expected to be luminosity-invariant,
i.e. it is a ``dusty wall''.

(iv) We have also considered the response function of a reprocessing
accretion disc, which predicts $\tau \propto \lambda^{4/3}$. We find
that, whereas the mean reverberation radii in the near-IR are
consistent with $\sim 70$~light-days, the delay map is strongly skewed
toward small values and peaks at radii consistent with dust in the
accretion disc. The existence of this dust, which implies that the
accretion disc extends out to relatively large radii (of \mbox{$R_{\rm
    out} \sim 1700-3000\,r_{\rm g}$}), is a prerequisite for the
recent models of \citet{Czerny17} and \citet{Baskin18}. These models
explain both the broad emission line region and the dusty torus as
part of the same dusty outflow launched by radiation pressure from the
outer regions of the accretion disc. It would be desirable to follow
up on this result with a near-IR reverberation campaign of a higher
cadence than ours.

\vspace*{0.2cm}

In the future, we plan to use our near-IR spectroscopy of NGC~5548 to
also study the coronal emission lines and the profile variations of
the broad emission lines, both of which are expected to further
constrain torus models. In general, our spectroscopic dust
reverberation programme currently includes four more sources (Mrk~876,
Mrk~110, Mrk~509 and 3C~273) and the results from these campaigns are
forthcoming.

\section*{Acknowledgments}

H.L. thanks Nancy Levenson for supporting this project in its initial
stages and Chris Done for the many enlightening conversions on the
physics of accretion discs. H.L., M.J.W. and D.K. acknowledge the
Science and Technology Facilities Council (STFC) through grant
ST/P000541/1 for support. C.P. acknowledges support from the National
Science Foundation (NSF) grant no. 1616828. G.J.F. acknowledges
support by NSF (1816537), NASA (ATP 17-ATP17-0141), and STScI (HST-AR-
15018). K.H. acknowledges support from STFC grant ST/R000824/1.

\bibliography{/Users/herminelandt/references}

\begin{thebibliography}{}

\bibitem[\protect\citeauthoryear{Abazajian et~al.}{Abazajian
  et~al.}{2009}]{SloanDR7}
Abazajian, K.~N., et~al. 2009, ApJS, 182, 543

\bibitem[\protect\citeauthoryear{Antonucci}{Antonucci}{1993}]{Ant93}
Antonucci, R. 1993, ARA\&A, 31, 473

\bibitem[\protect\citeauthoryear{Antonucci \& Miller}{Antonucci \&
  Miller}{1985}]{Ant85a}
Antonucci, R.,  \& Miller, J.~S. 1985, ApJ, 297, 621

\bibitem[\protect\citeauthoryear{{Barvainis}}{{Barvainis}}{1992}]{Barv92}
{Barvainis}, R. 1992, \apj, 400, 502

\bibitem[\protect\citeauthoryear{{Baskin} \& {Laor}}{{Baskin} \&
  {Laor}}{2018}]{Baskin18}
{Baskin}, A.,  \& {Laor}, A. 2018, \mnras, 474, 1970

\bibitem[\protect\citeauthoryear{{Bentz} \& {Katz}}{{Bentz} \&
  {Katz}}{2015}]{Bentz15}
{Bentz}, M.~C.,  \& {Katz}, S. 2015, \pasp, 127, 67

\bibitem[\protect\citeauthoryear{Bentz et~al.}{Bentz et~al.}{2009}]{Bentz09}
Bentz, M.~C., Peterson, B.~M., Netzer, H., Pogge, R.~W.,  \& Vestergaard, M.
  2009, ApJ, 697, 160

\bibitem[\protect\citeauthoryear{{Bentz} et~al.}{{Bentz}
  et~al.}{2008}]{Bentz08}
{Bentz}, M.~C., et~al. 2008, \apjl, 689, L21

\bibitem[\protect\citeauthoryear{{Blandford} \& {McKee}}{{Blandford} \&
  {McKee}}{1982}]{Bla82b}
{Blandford}, R.~D.,  \& {McKee}, C.~F. 1982, \apj, 255, 419

\bibitem[\protect\citeauthoryear{{Brown} et~al.}{{Brown} et~al.}{2013}]{LCOGT}
{Brown}, T.~M., et~al. 2013, \pasp, 125, 1031

\bibitem[\protect\citeauthoryear{{Burtscher} et~al.}{{Burtscher}
  et~al.}{2013}]{Bur13}
{Burtscher}, L., et~al. 2013, \aap, 558, A149

\bibitem[\protect\citeauthoryear{{Clavel}, {Wamsteker}, \& {Glass}}{{Clavel}
  et~al.}{1989}]{Clavel89}
{Clavel}, J., {Wamsteker}, W.,  \& {Glass}, I.~S. 1989, \apj, 337, 236

\bibitem[\protect\citeauthoryear{Cohen et~al.}{Cohen et~al.}{1999}]{Coh99}
Cohen, M.~H., Ogle, P.~M., Tran, H.~D., Goodrich, R.~W.,  \& Miller, J.~S.
  1999, AJ, 118, 1963

\bibitem[\protect\citeauthoryear{{Collier} et~al.}{{Collier}
  et~al.}{1999}]{Collier99}
{Collier}, S., {Horne}, K., {Wanders}, I.,  \& {Peterson}, B.~M. 1999, \mnras,
  302, L24

\bibitem[\protect\citeauthoryear{{Collins} et~al.}{{Collins}
  et~al.}{2017}]{astroimagej}
{Collins}, K.~A., {Kielkopf}, J.~F., {Stassun}, K.~G.,  \& {Hessman}, F.~V.
  2017, \aj, 153, 77

\bibitem[\protect\citeauthoryear{Cushing, Vacca, \& Rayner}{Cushing
  et~al.}{2004}]{Cush04}
Cushing, M.~C., Vacca, W.~D.,  \& Rayner, J.~T. 2004, PASP, 116, 362

\bibitem[\protect\citeauthoryear{{Czerny} et~al.}{{Czerny}
  et~al.}{2016}]{Czerny16}
{Czerny}, B., {Du}, P., {Wang}, J.-M.,  \& {Karas}, V. 2016, \apj, 832, 15

\bibitem[\protect\citeauthoryear{{Czerny} \& {Hryniewicz}}{{Czerny} \&
  {Hryniewicz}}{2011}]{Czerny11}
{Czerny}, B.,  \& {Hryniewicz}, K. 2011, \aap, 525, L8

\bibitem[\protect\citeauthoryear{{Czerny} et~al.}{{Czerny}
  et~al.}{2017}]{Czerny17}
{Czerny}, B., et~al. 2017, \apj, 846, 154

\bibitem[\protect\citeauthoryear{{De Rosa} et~al.}{{De Rosa}
  et~al.}{2015}]{Storm1}
{De Rosa}, G., et~al. 2015, \apj, 806, 128

\bibitem[\protect\citeauthoryear{{Denney} et~al.}{{Denney}
  et~al.}{2010}]{Denney10}
{Denney}, K.~D., et~al. 2010, \apj, 721, 715

\bibitem[\protect\citeauthoryear{Dickey \& Lockman}{Dickey \&
  Lockman}{1990}]{DL90}
Dickey, J.~M.,  \& Lockman, F.~J. 1990, ARA\&A, 28, 215

\bibitem[\protect\citeauthoryear{{Draine}}{{Draine}}{2016}]{Draine16}
{Draine}, B.~T. 2016, \apj, 831, 109

\bibitem[\protect\citeauthoryear{{Draine} \& {Hensley}}{{Draine} \&
  {Hensley}}{2017}]{Draine17}
{Draine}, B.~T.,  \& {Hensley}, B.~S. 2017, arXiv e-prints

\bibitem[\protect\citeauthoryear{{Edelson} et~al.}{{Edelson}
  et~al.}{2015}]{Storm2}
{Edelson}, R., et~al. 2015, \apj, 806, 129

\bibitem[\protect\citeauthoryear{{Elitzur} \& {Shlosman}}{{Elitzur} \&
  {Shlosman}}{2006}]{Elitzur06}
{Elitzur}, M.,  \& {Shlosman}, I. 2006, \apjl, 648, L101

\bibitem[\protect\citeauthoryear{{Fausnaugh}}{{Fausnaugh}}{2017}]{Faus17}
{Fausnaugh}, M.~M. 2017, \pasp, 129, 024007

\bibitem[\protect\citeauthoryear{{Fausnaugh} et~al.}{{Fausnaugh}
  et~al.}{2016}]{Storm3}
{Fausnaugh}, M.~M., et~al. 2016, \apj, 821, 56

\bibitem[\protect\citeauthoryear{{Ferguson}, {Korista}, \&
  {Ferland}}{{Ferguson} et~al.}{1997}]{Ferg97}
{Ferguson}, J.~W., {Korista}, K.~T.,  \& {Ferland}, G.~J. 1997, \apjs, 110, 287

\bibitem[\protect\citeauthoryear{Frank, King, \& Raine}{Frank
  et~al.}{2002}]{FKR}
Frank, J., King, A.,  \& Raine, D. 2002, Accretion Power in Astrophysics
  (Cambridge University Press)

\bibitem[\protect\citeauthoryear{{Fuller} et~al.}{{Fuller}
  et~al.}{2019}]{Fuller19}
{Fuller}, L., et~al. 2019, \mnras, 483, 3404

\bibitem[\protect\citeauthoryear{{Garc{\'{\i}}a-Burillo}
  et~al.}{{Garc{\'{\i}}a-Burillo} et~al.}{2016}]{Garcia16}
{Garc{\'{\i}}a-Burillo}, S., et~al. 2016, \apjl, 823, L12

\bibitem[\protect\citeauthoryear{{Gardner} \& {Done}}{{Gardner} \&
  {Done}}{2017}]{Gardner17}
{Gardner}, E.,  \& {Done}, C. 2017, \mnras, 470, 3591

\bibitem[\protect\citeauthoryear{{Gaskell} \& {Peterson}}{{Gaskell} \&
  {Peterson}}{1987}]{Gas87}
{Gaskell}, C.~M.,  \& {Peterson}, B.~M. 1987, \apjs, 65, 1

\bibitem[\protect\citeauthoryear{{Glass}}{{Glass}}{2004}]{Glass04}
{Glass}, I.~S. 2004, \mnras, 350, 1049

\bibitem[\protect\citeauthoryear{{Goad} et~al.}{{Goad} et~al.}{2016}]{Storm4}
{Goad}, M.~R., et~al. 2016, \apj, 824, 11

\bibitem[\protect\citeauthoryear{{Goad}, {Korista}, \& {Ruff}}{{Goad}
  et~al.}{2012}]{Goad12}
{Goad}, M.~R., {Korista}, K.~T.,  \& {Ruff}, A.~J. 2012, \mnras, 426, 3086

\bibitem[\protect\citeauthoryear{{Greiner} et~al.}{{Greiner}
  et~al.}{2008}]{grond}
{Greiner}, J., et~al. 2008, \pasp, 120, 405

\bibitem[\protect\citeauthoryear{{Grier} et~al.}{{Grier}
  et~al.}{2013}]{Grier13b}
{Grier}, C.~J., et~al. 2013, \apj, 773, 90

\bibitem[\protect\citeauthoryear{{H{\"o}nig} et~al.}{{H{\"o}nig}
  et~al.}{2013}]{Hoenig13}
{H{\"o}nig}, S.~F., et~al. 2013, \apj, 771, 87

\bibitem[\protect\citeauthoryear{{Horne}}{{Horne}}{1986}]{Horne86a}
{Horne}, K. 1986, \pasp, 98, 609

\bibitem[\protect\citeauthoryear{{Horne}}{{Horne}}{1994}]{Horne94}
{Horne}, K. 1994, in Astronomical Society of the Pacific Conference Series,
  Vol.~69, Reverberation Mapping of the Broad-Line Region in Active Galactic
  Nuclei, ed. P.~M. {Gondhalekar}, K.~{Horne}, \& B.~M. {Peterson}, 23

\bibitem[\protect\citeauthoryear{{Horne}, {Welsh}, \& {Peterson}}{{Horne}
  et~al.}{1991}]{Horne91}
{Horne}, K., {Welsh}, W.~F.,  \& {Peterson}, B.~M. 1991, \apjl, 367, L5

\bibitem[\protect\citeauthoryear{{Hubeny} et~al.}{{Hubeny}
  et~al.}{2000}]{Hubeny00}
{Hubeny}, I., {Agol}, E., {Blaes}, O.,  \& {Krolik}, J.~H. 2000, \apj, 533, 710

\bibitem[\protect\citeauthoryear{{Imanishi} et~al.}{{Imanishi}
  et~al.}{2018}]{Imanishi18}
{Imanishi}, M., {Nakanishi}, K., {Izumi}, T.,  \& {Wada}, K. 2018, \apjl, 853,
  L25

\bibitem[\protect\citeauthoryear{{Jones}}{{Jones}}{2004}]{Jones04}
{Jones}, A.~P. 2004, in Astronomical Society of the Pacific Conference Series,
  Vol. 309, Astrophysics of Dust, ed. A.~N. {Witt}, G.~C. {Clayton}, \& B.~T.
  {Draine}, 347

\bibitem[\protect\citeauthoryear{{Kaastra} et~al.}{{Kaastra}
  et~al.}{2014}]{Kaastra14}
{Kaastra}, J.~S., et~al. 2014, Science, 345, 64

\bibitem[\protect\citeauthoryear{{Kawaguchi} \& {Mori}}{{Kawaguchi} \&
  {Mori}}{2010}]{Kaw10}
{Kawaguchi}, T.,  \& {Mori}, M. 2010, \apjl, 724, L183

\bibitem[\protect\citeauthoryear{{Kawaguchi} \& {Mori}}{{Kawaguchi} \&
  {Mori}}{2011}]{Kaw11}
{Kawaguchi}, T.,  \& {Mori}, M. 2011, \apj, 737, 105

\bibitem[\protect\citeauthoryear{{Keating} et~al.}{{Keating}
  et~al.}{2012}]{Keating12}
{Keating}, S.~K., {Everett}, J.~E., {Gallagher}, S.~C.,  \& {Deo}, R.~P. 2012,
  \apj, 749, 32

\bibitem[\protect\citeauthoryear{{Kishimoto} et~al.}{{Kishimoto}
  et~al.}{2008}]{Kish08}
{Kishimoto}, M., {Antonucci}, R., {Blaes}, O., {Lawrence}, A., {Boisson}, C.,
  {Albrecht}, M.,  \& {Leipski}, C. 2008, \nat, 454, 492

\bibitem[\protect\citeauthoryear{{Kishimoto} et~al.}{{Kishimoto}
  et~al.}{2011}]{Kish11}
{Kishimoto}, M., {H{\"o}nig}, S.~F., {Antonucci}, R., {Barvainis}, R.,
  {Kotani}, T., {Tristram}, K.~R.~W., {Weigelt}, G.,  \& {Levin}, K. 2011,
  \aap, 527, A121

\bibitem[\protect\citeauthoryear{{Kishimoto} et~al.}{{Kishimoto}
  et~al.}{2009}]{Kish09}
{Kishimoto}, M., {H{\"o}nig}, S.~F., {Antonucci}, R., {Kotani}, T.,
  {Barvainis}, R., {Tristram}, K.~R.~W.,  \& {Weigelt}, G. 2009, \aap, 507, L57

\bibitem[\protect\citeauthoryear{{Kishimoto} et~al.}{{Kishimoto}
  et~al.}{2013}]{Kish13}
{Kishimoto}, M., et~al. 2013, \apjl, 775, L36

\bibitem[\protect\citeauthoryear{{Kishimoto} et~al.}{{Kishimoto}
  et~al.}{2007}]{Kish07}
{Kishimoto}, M., {H{\"o}nig}, S.~F., {Beckert}, T.,  \& {Weigelt}, G. 2007,
  \aap, 476, 713

\bibitem[\protect\citeauthoryear{{K\"onigl} \& {Kartje}}{{K\"onigl} \&
  {Kartje}}{1994}]{Koenigl94}
{K\"onigl}, A.,  \& {Kartje}, J.~F. 1994, \apj, 434, 446

\bibitem[\protect\citeauthoryear{Koratkar \& Blaes}{Koratkar \&
  Blaes}{1999}]{Korat99}
Koratkar, A.,  \& Blaes, O. 1999, PASP, 111, 1

\bibitem[\protect\citeauthoryear{{Koshida} et~al.}{{Koshida}
  et~al.}{2014}]{Kosh14}
{Koshida}, S., et~al. 2014, \apj, 788, 159

\bibitem[\protect\citeauthoryear{{Koshida} et~al.}{{Koshida}
  et~al.}{2009}]{Kosh09}
{Koshida}, S., et~al. 2009, \apjl, 700, L109

\bibitem[\protect\citeauthoryear{{Kraemer} et~al.}{{Kraemer}
  et~al.}{1998}]{Kraemer98}
{Kraemer}, S.~B., {Crenshaw}, D.~M., {Filippenko}, A.~V.,  \& {Peterson}, B.~M.
  1998, \apj, 499, 719

\bibitem[\protect\citeauthoryear{{Krolik}}{{Krolik}}{2007}]{Kro07}
{Krolik}, J.~H. 2007, \apj, 661, 52

\bibitem[\protect\citeauthoryear{{Krolik} \& {Begelman}}{{Krolik} \&
  {Begelman}}{1986}]{Kro86}
{Krolik}, J.~H.,  \& {Begelman}, M.~C. 1986, \apjl, 308, L55

\bibitem[\protect\citeauthoryear{Krolik \& Begelman}{Krolik \&
  Begelman}{1988}]{Kro88}
Krolik, J.~H.,  \& Begelman, M.~C. 1988, ApJ, 329, 702

\bibitem[\protect\citeauthoryear{Landt et~al.}{Landt et~al.}{2011a}]{L11b}
Landt, H., Bentz, M.~C., Peterson, B.~M., Elvis, M., Ward, M.~J., Korista,
  K.~T.,  \& Karovska, M. 2011a, MNRAS, 413, L106

\bibitem[\protect\citeauthoryear{Landt et~al.}{Landt et~al.}{2008}]{L08a}
Landt, H., Bentz, M.~C., Ward, M.~J., Elvis, M., Peterson, B.~M., Korista,
  K.~T.,  \& Karovska, M. 2008, ApJS, 174, 282

\bibitem[\protect\citeauthoryear{Landt, Buchanan, \& Barmby}{Landt
  et~al.}{2010}]{L10b}
Landt, H., Buchanan, C.~L.,  \& Barmby, P. 2010, MNRAS, 408, 1982

\bibitem[\protect\citeauthoryear{Landt et~al.}{Landt et~al.}{2011b}]{L11a}
Landt, H., Elvis, M., Ward, M.~J., Bentz, M.~C., Korista, K.~T.,  \& Karovska,
  M. 2011b, MNRAS, 414, 218

\bibitem[\protect\citeauthoryear{{Landt} et~al.}{{Landt} et~al.}{2014}]{L14}
{Landt}, H., {Ward}, M.~J., {Elvis}, M.,  \& {Karovska}, M. 2014, \mnras, 439,
  1051

\bibitem[\protect\citeauthoryear{Landt et~al.}{Landt et~al.}{2015}]{L15b}
Landt, H., Ward, M.~J., Steenbrugge, K.~C.,  \& Ferland, G.~J. 2015, MNRAS,
  454, 3688

\bibitem[\protect\citeauthoryear{{Laor} \& {Draine}}{{Laor} \&
  {Draine}}{1993}]{Laor93}
{Laor}, A.,  \& {Draine}, B.~T. 1993, \apj, 402, 441

\bibitem[\protect\citeauthoryear{{Lawrence}}{{Lawrence}}{1987}]{Law87}
{Lawrence}, A. 1987, \pasp, 99, 309

\bibitem[\protect\citeauthoryear{{Lawrence}}{{Lawrence}}{1991}]{Law91}
{Lawrence}, A. 1991, \mnras, 252, 586

\bibitem[\protect\citeauthoryear{{Lodders}}{{Lodders}}{2003}]{Lod03}
{Lodders}, K. 2003, \apj, 591, 1220

\bibitem[\protect\citeauthoryear{{Lopez-Rodriguez} et~al.}{{Lopez-Rodriguez}
  et~al.}{2017}]{Lopez17}
{Lopez-Rodriguez}, E., et~al. 2017, \mnras, 464, 1762

\bibitem[\protect\citeauthoryear{{Lopez-Rodriguez} et~al.}{{Lopez-Rodriguez}
  et~al.}{2013}]{Lopez13}
{Lopez-Rodriguez}, E., et~al. 2013, \mnras, 431, 2723

\bibitem[\protect\citeauthoryear{Lumsden et~al.}{Lumsden et~al.}{2001}]{Lum01}
Lumsden, S.~L., Heisler, C.~A., Bailey, J.~A., Hough, J.~H.,  \& Young, S.
  2001, MNRAS, 327, 459

\bibitem[\protect\citeauthoryear{{Maiolino}, {Marconi}, \& {Oliva}}{{Maiolino}
  et~al.}{2001}]{Mai01b}
{Maiolino}, R., {Marconi}, A.,  \& {Oliva}, E. 2001, \aap, 365, 37

\bibitem[\protect\citeauthoryear{{Markwardt}}{{Markwardt}}{2009}]{mpfit}
{Markwardt}, C.~B. 2009, in ASP Conference Series, Vol. 411, Astronomical Data
  Analysis Software and Systems XVIII, ed. D.~Bohlender, P.~Dowler, \&
  D.~Durand (San Francisco: Astronomical Society of the Pacific),
  arXiv:0902.2850v1

\bibitem[\protect\citeauthoryear{{Mathur} et~al.}{{Mathur}
  et~al.}{2017}]{Storm7}
{Mathur}, S., et~al. 2017, \apj, 846, 55

\bibitem[\protect\citeauthoryear{{McGregor} et~al.}{{McGregor}
  et~al.}{2003}]{nifs}
{McGregor}, P.~J., et~al. 2003, in Proc. SPIE, Vol. 4841, Instrument Design and
  Performance for Optical/Infrared Ground-based Telescopes, ed. M.~{Iye} \&
  A.~F.~M. {Moorwood}, 1581

\bibitem[\protect\citeauthoryear{{Mehdipour} et~al.}{{Mehdipour}
  et~al.}{2015}]{Meh15a}
{Mehdipour}, M., et~al. 2015, \aap, 575, A22

\bibitem[\protect\citeauthoryear{{Minezaki} et~al.}{{Minezaki}
  et~al.}{2004}]{Min04}
{Minezaki}, T., {Yoshii}, Y., {Kobayashi}, Y., {Enya}, K., {Suganuma}, M.,
  {Tomita}, H., {Aoki}, T.,  \& {Peterson}, B.~A. 2004, \apjl, 600, L35

\bibitem[\protect\citeauthoryear{{Mor} \& {Netzer}}{{Mor} \&
  {Netzer}}{2012}]{Mor12}
{Mor}, R.,  \& {Netzer}, H. 2012, \mnras, 420, 526

\bibitem[\protect\citeauthoryear{{Morganson} et~al.}{{Morganson}
  et~al.}{2012}]{Morg12}
{Morganson}, E., et~al. 2012, \aj, 143, 142

\bibitem[\protect\citeauthoryear{{Nelson}}{{Nelson}}{1996}]{Nel96}
{Nelson}, B.~O. 1996, \apjl, 465, L87

\bibitem[\protect\citeauthoryear{Nenkova et~al.}{Nenkova et~al.}{2008}]{Nen08a}
Nenkova, M., Sirocky, M.~M., Ivezi\'c, Z.,  \& Elitzur, M. 2008, ApJ, 685, 147

\bibitem[\protect\citeauthoryear{{Netzer}}{{Netzer}}{1987}]{Netzer87}
{Netzer}, H. 1987, \mnras, 225, 55

\bibitem[\protect\citeauthoryear{{Netzer}}{{Netzer}}{2015}]{Netzer15}
{Netzer}, H. 2015, \araa, 53, 365

\bibitem[\protect\citeauthoryear{{Netzer} \& {Laor}}{{Netzer} \&
  {Laor}}{1993}]{Netzer93a}
{Netzer}, H.,  \& {Laor}, A. 1993, \apjl, 404, L51

\bibitem[\protect\citeauthoryear{{Oknyanskij} \& {Horne}}{{Oknyanskij} \&
  {Horne}}{2001}]{Okn01}
{Oknyanskij}, V.~L.,  \& {Horne}, K. 2001, in Astronomical Society of the
  Pacific Conference Series, Vol. 224, Probing the Physics of Active Galactic
  Nuclei, ed. B.~M. {Peterson}, R.~W. {Pogge}, \& R.~S. {Polidan}, 149

\bibitem[\protect\citeauthoryear{{Packham} et~al.}{{Packham}
  et~al.}{2005}]{Pack05}
{Packham}, C., {Radomski}, J.~T., {Roche}, P.~F., {Aitken}, D.~K., {Perlman},
  E., {Alonso-Herrero}, A., {Colina}, L.,  \& {Telesco}, C.~M. 2005, \apjl,
  618, L17

\bibitem[\protect\citeauthoryear{{Pancoast}, {Brewer}, \& {Treu}}{{Pancoast}
  et~al.}{2011}]{Pan11}
{Pancoast}, A., {Brewer}, B.~J.,  \& {Treu}, T. 2011, \apj, 730, 139

\bibitem[\protect\citeauthoryear{{Pancoast} et~al.}{{Pancoast}
  et~al.}{2014}]{Pan14}
{Pancoast}, A., {Brewer}, B.~J., {Treu}, T., {Park}, D., {Barth}, A.~J.,
  {Bentz}, M.~C.,  \& {Woo}, J.-H. 2014, \mnras, 445, 3073

\bibitem[\protect\citeauthoryear{{Pei} et~al.}{{Pei} et~al.}{2017}]{Storm5}
{Pei}, L., et~al. 2017, \apj, 837, 131

\bibitem[\protect\citeauthoryear{{Peng} et~al.}{{Peng} et~al.}{2002}]{galfit}
{Peng}, C.~Y., {Ho}, L.~C., {Impey}, C.~D.,  \& {Rix}, H.-W. 2002, \aj, 124,
  266

\bibitem[\protect\citeauthoryear{Peterson}{Peterson}{1993}]{Pet93}
Peterson, B.~M. 1993, PASP, 105, 247

\bibitem[\protect\citeauthoryear{Peterson}{Peterson}{1997}]{Peterson}
Peterson, B.~M. 1997, An Introduction to Active Galactic Nuclei (Cambridge
  University Press)

\bibitem[\protect\citeauthoryear{{Peterson} et~al.}{{Peterson}
  et~al.}{2013}]{Pet13}
{Peterson}, B.~M., et~al. 2013, \apj, 779, 109

\bibitem[\protect\citeauthoryear{Peterson et~al.}{Peterson
  et~al.}{2004}]{Pet04}
Peterson, B.~M., et~al. 2004, ApJ, 613, 682

\bibitem[\protect\citeauthoryear{{Pier} \& {Krolik}}{{Pier} \&
  {Krolik}}{1992}]{Pier92a}
{Pier}, E.~A.,  \& {Krolik}, J.~H. 1992, \apjl, 399, L23

\bibitem[\protect\citeauthoryear{{Pier} \& {Voit}}{{Pier} \&
  {Voit}}{1995}]{Pier95}
{Pier}, E.~A.,  \& {Voit}, G.~M. 1995, \apj, 450, 628

\bibitem[\protect\citeauthoryear{{Pott} et~al.}{{Pott} et~al.}{2010}]{Pott10}
{Pott}, J.-U., {Malkan}, M.~A., {Elitzur}, M., {Ghez}, A.~M., {Herbst}, T.~M.,
  {Sch{\"o}del}, R.,  \& {Woillez}, J. 2010, \apj, 715, 736

\bibitem[\protect\citeauthoryear{{Ramos Almeida} et~al.}{{Ramos Almeida}
  et~al.}{2011}]{Ramos11}
{Ramos Almeida}, C., et~al. 2011, \apj, 731, 92

\bibitem[\protect\citeauthoryear{Rayner et~al.}{Rayner et~al.}{2003}]{Ray03}
Rayner, J.~T., Toomey, D.~W., Onaka, P.~M., Denault, A.~J., Stahlberger, W.~E.,
  Vacca, W.~D., Cushing, M.~C.,  \& Wang, S. 2003, PASP, 115, 362

\bibitem[\protect\citeauthoryear{{Rees} et~al.}{{Rees} et~al.}{1969}]{Rees69}
{Rees}, M.~J., {Silk}, J.~I., {Werner}, M.~W.,  \& {Wickramasinghe}, N.~C.
  1969, \nat, 223, 788

\bibitem[\protect\citeauthoryear{{Salpeter}}{{Salpeter}}{1977}]{Sal77}
{Salpeter}, E.~E. 1977, \araa, 15, 267

\bibitem[\protect\citeauthoryear{{Schn{\"u}lle} et~al.}{{Schn{\"u}lle}
  et~al.}{2013}]{Schnuelle13}
{Schn{\"u}lle}, K., {Pott}, J.-U., {Rix}, H.-W., {Decarli}, R., {Peterson},
  B.~M.,  \& {Vacca}, W. 2013, \aap, 557, L13

\bibitem[\protect\citeauthoryear{{Schn{\"u}lle} et~al.}{{Schn{\"u}lle}
  et~al.}{2015}]{Schnuelle15}
{Schn{\"u}lle}, K., {Pott}, J.-U., {Rix}, H.-W., {Peterson}, B.~M., {De Rosa},
  G.,  \& {Shappee}, B. 2015, \aap, 578, A57

\bibitem[\protect\citeauthoryear{{Sch{\"o}nell} et~al.}{{Sch{\"o}nell}
  et~al.}{2017}]{Schoenell17}
{Sch{\"o}nell}, A.~J., Jr., {Storchi-Bergmann}, T., {Riffel}, R.~A.,  \&
  {Riffel}, R. 2017, \mnras, 464, 1771

\bibitem[\protect\citeauthoryear{{Shen} et~al.}{{Shen} et~al.}{2015}]{Shen15}
{Shen}, Y., et~al. 2015, \apjs, 216, 4

\bibitem[\protect\citeauthoryear{{Shen} et~al.}{{Shen} et~al.}{2016}]{Shen16}
{Shen}, Y., et~al. 2016, \apj, 818, 30

\bibitem[\protect\citeauthoryear{{Sitko} et~al.}{{Sitko} et~al.}{1993}]{Sit93}
{Sitko}, M.~L., {Sitko}, A.~K., {Siemiginowska}, A.,  \& {Szczerba}, R. 1993,
  \apj, 409, 139

\bibitem[\protect\citeauthoryear{Skrutskie et~al.}{Skrutskie
  et~al.}{2006}]{2MASS}
Skrutskie, M.~F., et~al. 2006, AJ, 131, 1163

\bibitem[\protect\citeauthoryear{{Starkey} et~al.}{{Starkey}
  et~al.}{2017}]{Storm6}
{Starkey}, D., et~al. 2017, \apj, 835, 65

\bibitem[\protect\citeauthoryear{{Starkey}, {Horne}, \& {Villforth}}{{Starkey}
  et~al.}{2016}]{Starkey16}
{Starkey}, D.~A., {Horne}, K.,  \& {Villforth}, C. 2016, \mnras, 456, 1960

\bibitem[\protect\citeauthoryear{{Storchi-Bergmann} et~al.}{{Storchi-Bergmann}
  et~al.}{2009}]{Storchi09}
{Storchi-Bergmann}, T., {McGregor}, P.~J., {Riffel}, R.~A., {Sim{\~o}es Lopes},
  R., {Beck}, T.,  \& {Dopita}, M. 2009, \mnras, 394, 1148

\bibitem[\protect\citeauthoryear{{Storchi-Bergmann} et~al.}{{Storchi-Bergmann}
  et~al.}{2017}]{Storchi17}
{Storchi-Bergmann}, T., {Schimoia}, J.~S., {Peterson}, B.~M., {Elvis}, M.,
  {Denney}, K.~D., {Eracleous}, M.,  \& {Nemmen}, R.~S. 2017, \apj, 835, 236

\bibitem[\protect\citeauthoryear{{Suganuma} et~al.}{{Suganuma}
  et~al.}{2006}]{Sug06}
{Suganuma}, M., et~al. 2006, \apj, 639, 46

\bibitem[\protect\citeauthoryear{{Tran}}{{Tran}}{2003}]{Tran03}
{Tran}, H.~D. 2003, \apj, 583, 632

\bibitem[\protect\citeauthoryear{{Tristram} et~al.}{{Tristram}
  et~al.}{2009}]{Tri09}
{Tristram}, K.~R.~W., et~al. 2009, \aap, 502, 67

\bibitem[\protect\citeauthoryear{Urry \& Padovani}{Urry \&
  Padovani}{1995}]{Urry95}
Urry, C.~M.,  \& Padovani, P. 1995, PASP, 107, 803

\bibitem[\protect\citeauthoryear{{van Groningen} \& {Wanders}}{{van Groningen}
  \& {Wanders}}{1992}]{vanGron92}
{van Groningen}, E.,  \& {Wanders}, I. 1992, \pasp, 104, 700

\bibitem[\protect\citeauthoryear{{Vazquez} et~al.}{{Vazquez}
  et~al.}{2015}]{Vazquez15}
{Vazquez}, B., et~al. 2015, \apj, 801, 127

\bibitem[\protect\citeauthoryear{{White} \& {Peterson}}{{White} \&
  {Peterson}}{1994}]{White94}
{White}, R.~J.,  \& {Peterson}, B.~M. 1994, \pasp, 106, 879

\bibitem[\protect\citeauthoryear{{Zu}, {Kochanek}, \& {Peterson}}{{Zu}
  et~al.}{2011}]{Zu11}
{Zu}, Y., {Kochanek}, C.~S.,  \& {Peterson}, B.~M. 2011, \apj, 735, 80

\end{thebibliography}

\bsp
\label{lastpage}

\end{document}